\documentclass[preprint,12pt]{elsarticle}

\usepackage{graphicx}
\usepackage{amssymb}
\usepackage{amsmath}
\usepackage{commath}
\usepackage{mathtools}

\usepackage{lineno}
\usepackage{subfigure}
\usepackage{bm}
\usepackage{cleveref}

\usepackage{graphicx}
\usepackage{booktabs}
\usepackage{siunitx}

\usepackage[final]{changes}
\usepackage{multirow}

\usepackage{tikz}
\usetikzlibrary{shapes.geometric, arrows}

\tikzstyle{startstop} = [rectangle, rounded corners, minimum width=3cm, minimum height=1cm,text centered, draw=black, fill=red!30]
\tikzstyle{io} = [trapezium, trapezium left angle=70, trapezium right angle=110, minimum width=3cm, minimum height=1cm, text centered, draw=black, fill=blue!30]
\tikzstyle{process} = [rectangle, minimum width=3cm, minimum height=1cm, text centered, draw=black, fill=orange!30]
\tikzstyle{decision} = [diamond, minimum width=3cm, minimum height=1cm, text centered, draw=black, fill=green!30]
\tikzstyle{arrow} = [thick,->,>=stealth]

\biboptions{sort&compress}

\begin{document}

\begin{frontmatter}


\title{Coupled Metaball Discrete Element Lattice Boltzmann Method for Fluid-Particle Systems with non-spherical particle shapes: A sharp interface coupling scheme}


\author[wu]{Pei Zhang}
\author[wu]{Ling Qiu}
\author[changwei]{Yilin Chen}
\author[uq1]{A. Scheuermann}
\author[wu]{Ling Li}
\author[wu]{S.A. Galindo-Torres\corref{cor1}}
\cortext[cor1]{s.Torres@westlake.edu.cn}

\address[wu]{Key Laboratory of Coastal Environment and Resources of Zhejiang Province, School of Engineering, Westlake University, Hangzhou, China\fnref{wu}}
\address[changwei]{Changjiang Institute of Survey, Planning, Design and Research, Wuhan, China}
\address[uq1]{School of Civil Engineering, University of Queensland, Queensland, Australia\fnref{uq1}}

\begin{abstract}
Fluid-particle systems are very common in many natural processes and engineering applications. However, accurately and efficiently modelling fluid-particle systems with complex particle shapes is still a challenging task. Here, we present a numerical model that combines the advantages of Lattice Boltzmann Method (LBM) in solving complex flow problems and the capability of the recently introduced Metaball Discrete Element Method (MDEM) in handling non-spherical particle shapes. A sharp interface coupling scheme is developed and the numerical instability issues due to the discontinuity of interfaces are carefully addressed. A local refilling algorithm for new fluid nodes is proposed and special treatments are introduced to reduce numerical noises when two particles are close. The proposed model is validated by simulations of settling of a single sphere (with metaball representation) as well as a non-spherical particle in a viscous fluid. Good agreements are found comparing the simulations with experimental results which are also carried out in this study. The coupling scheme is also tested by multiple particle simulations which clearly illustrated the stability of the proposed model. Finally, numerical examples with complex particle shapes demonstrated that the proposed model can be a powerful tool for future applications such as shape-induced segregation in riverbeds, and phase transition of dense suspension.
\end{abstract}

\end{frontmatter}

%%podlozhnyuk2017efficient,cai2019diffusion,jing2017micromechanical,descantes2019classical, seelen2018granular
%%\linenumbers

\section{Introduction}\label{sec:intro}
\begin{figure}[b]
\centering     %%% not \center
\subfigure[Pebbles]{\label{fig:a}\includegraphics[width=60mm]{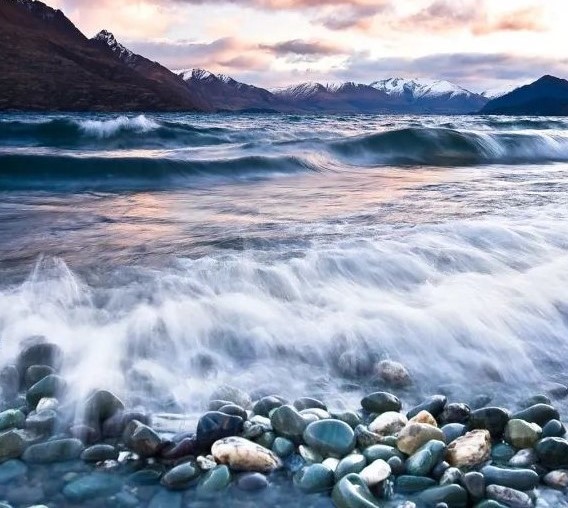}}
\subfigure[New Zealand’s Moeraki boulders]{\label{fig:b}\includegraphics[width=60mm]{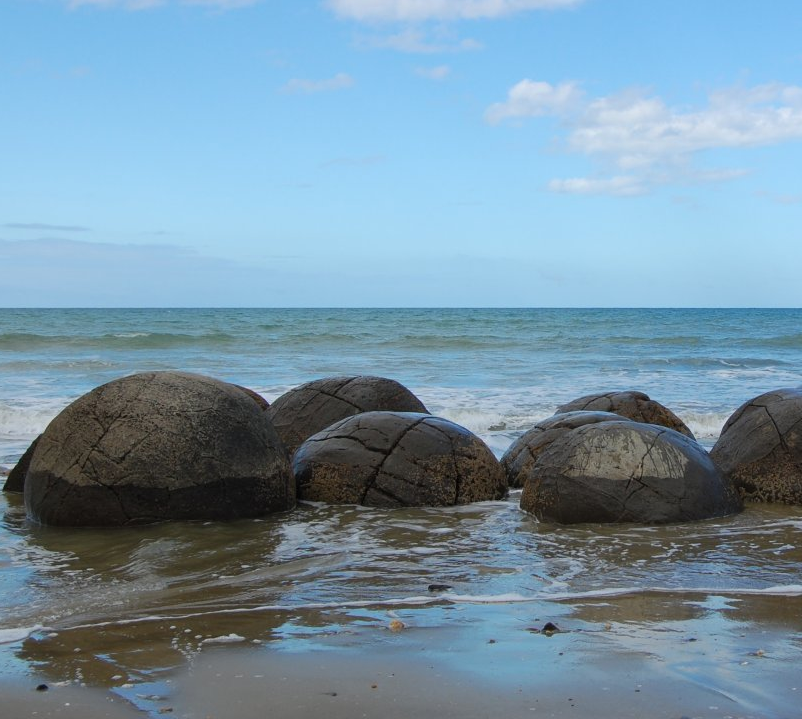}}
\caption{Some examples of general shaped particles with round features found in nature. (Image source: Google and Geological Society of Glasgow.)}
\label{fig:round_particle}
\end{figure}

Fluid-particle systems widely exist in both natural and industrial processes. Examples can be found from drug delivery within human bodies~\cite{faraji2009nanoparticles}, sediment transport~\cite{yuan2018water} in oceans and rivers, debris flows~\cite{trujillo2020smooth}, to particle mixing in a fluidized-bed reactor~\cite{derksen2014simulations}. One common feature of these systems is that the involved particles often have complex non-spherical shapes (Fig.~\ref{fig:round_particle}). The particle shape not only affects fluid-particle, particle-particle interactions at the individual particle level but also can dramatically influence the behaviours of fluid-particle systems at the macroscopic scale. For instance, the jamming of dense suspension is highly sensitive to particle shapes~\cite{guazzelli2018rheology}. Besides the importance of particle shapes for fluid-particle systems, it is still a challenging task to quantitatively describe the role of particle shapes and explain the underlay mechanisms. Difficulties arise from the fact that current experimental observations cannot always provide enough information, particularly about the mechanics occurring at the particle scale, due to measurement limitations. Therefore, numerical modellings become increasingly important for understanding fluid-particle systems. With increasing computational powers, particle scale resolved numerical methods become promising tools to explore details of flows and particle motions at both the microscopic and macroscopic scales.

One popular approach to simulate particle motions is the Discrete Element Method (DEM)~\cite{luding2008introduction}. DEM uses a bottom-up strategy where individual particle motions are tracked directly and particle-particle interactions are modelled at the particle scale. Classical DEM can only indirectly take into account the shape effects by rolling resistance since all particles are approximated as spheres~\cite{ai2011assessment}. Thus, many shape description methods are developed to address this issue. For example, complex shapes can be approximated by glueing spheres together~\cite{garcia2009clustered, ashmawy2003evaluating}, the collisions between particles are then simplified into sphere-sphere contacts. Although this approach is widely used, the sphere-clustering technique introduces an artificial surface roughness and a significant number of prime spheres are required to have a decent shape approximation~\cite{kruggel2008study}. Polyhedral DEM also draws a lot of attention due to its capability of describing complex shapes efficiently. However, it suffers from numerical instability due to the non-smooth nature of each polyhedron. The Sphero-polyhedron approach~\cite{alonso2008spheropolygons, alonso2009efficient, galindo2010minkowski,galindo2012breaking, galindo2009molecular} overcomes this issue by smoothing particle surfaces with a sphere. However, not every shape can be efficiently represented by a polyhedral mesh, a drawback that will be discussed further on. Particle shapes can also be described by distance functions such as the level set function. Level set DEM~\cite{kawamoto2016level} can handle particles with realistic shapes by directly cooperating with CT scan results. The main limitation is the high computational cost since each particle is represented by a points cloud. The recently developed metaball DEM~\cite{zhang2021metaball} describes particle shapes by a metaball function analytically. The contact between particles is modelled by solving an optimization problem. It shows great potential for handling non-spherical particles with rounded features without discretization of particle surfaces.

To solve fluid-particle interactions, DEM needs to be coupled with Computational Fluid Dynamics (CFD) methods. The Lattice Boltzmann Method (LBM) has emerged as an effective CFD solver during the last decades, and it has attracted enormous interest in simulating complex flows including fluid-particle systems~\cite{ladd1994numerical,zhang2016lattice, cui20122d,wang2013lattice, galindo2013coupled, zhang2017efficient}. LBM has several unique advantages that make it suitable to couple with DEM. First, LBM enjoys high parallelization efficiency due to the locality of the collision operator, where the computational cost is always a bottleneck of fully solved fluid-particle simulations. Furthermore, the kinetic nature of LBM ensures its capability in handling complex moving boundary conditions with simple algorithms. The DEM-LBM coupling schemes can be classified into two categories: diffuse interface approach and sharp interface approach. Diffuse interface schemes handle the discontinuity at solid-fluid boundaries by smoothing the interface. Most Immersed Boundary Methods (IBM)~\cite{peskin2002immersed, feng2004immersed, luo2007modified, niu2006momentum, wu2009implicit} belong to this category, where the influence of solid boundaries is replaced by a smoothed external force field. In the partially saturated cells method (PSM)~\cite{noble1998lattice, cook2004direct, feng2007coupled}, the solid boundaries are introduced by the solid volume fraction, therefore, the exact boundary position does not exist in the fluid solver. Although diffuse interface approaches benefit from smooth transitions between solid and fluid nodes and fewer fluctuations in hydrodynamic forces, the non-physical diffuse interface representation limits its accuracy. For instance, it is found that IBM can only achieve first-order accuracy when simulating porous media flows and PSM underestimates the permeability systematically~\cite{chen2020intercomparison}. On the other hand, the sharp interface approaches treat solid boundaries without smoothing. Within the LBM framework, it is straightforward to handle sharp interfaces by applying the bounce-back rule: fluid molecules that contact the solid surface are reflected back to the fluid domain with opposite velocity. The simple bounce-back scheme approximates interfaces as stairwise boundaries which may damage overall accuracy~\cite{ladd1994numerical}. Thus, Bouzidi et al.~\cite{bouzidi2001momentum} introduced an improved bounce-back scheme where the missing distribution functions are interpolated. It is further developed by Yu et al.~\cite{yu2003viscous} with a unified scheme by estimating the distribution at boundaries. It is found that the interpolated bounce-back schemes (IBB) have second-order accuracy in space~\cite{peng2016implementation, chen2020efficient}. The trade-off of sharp interface representations is less numerical stability: the hydrodynamic forces are considerable noisier than diffuse interface schemes~\cite{peng2016implementation}. Special treatments are also required for new fluid nodes due to moving boundaries~\cite{lallemand2003lattice, fang2002lattice, zhang2021coupled}. Therefore, diffuse interface approaches are widely used for DEM-LBM coupling despite sharp interface schemes having better accuracy in general~\cite{zhang2022random}. Recently, Peng et al.~\cite{peng2019comparative,peng2019comparative2} conducted comprehensive comparisons between IBM and IBB for both laminar and turbulence flows. It is shown that IBB is second-order accuracy for velocity, hydrodynamic force/torque, and stress, where diffuse interface IBM only hold first-order accuracy when simulating laminar flows~\cite{peng2019comparative}. Furthermore, IBM fails to correctly capture the velocity gradient within the diffuse interface for turbulent flows~\cite{peng2019comparative2}.

In terms of particle shape, the majority of DEM-LBM schemes use spheres with diffuse interface coupling schemes~\cite{feng2010combined, feng2007coupled, zhang2016lattice}. Galindo-Torres~\cite{galindo2013coupled} extended the DEM-LBM model for generally shaped particles (even non-convex ones) by using the sphero-polyhedron technique. Recently, Wang et al.~\cite{wang2021coupled} introduced a polygonal DEM-LBM model with an energy-conserving contact algorithm. Although these latest developments can handle fluid-particle interactions with complex shapes, it is still not an easy task to handle particles with round shapes such as river pebbles due to the large numbers of vertices required by the surface mesh.

The goal of this work is to provide a sharp interface coupling model that combines the efficiency of LBM in solving flows and the capability of MDEM in handling non-spherical particles to simulate fluid-particle systems. The structure of the paper is organized as follows: Sec.~\ref{sec:mathod} describes the basics of MDEM and LBM, the ideas, approximations and detailed implementations of the coupling scheme. The presented model is validated by comparing with settling of a sphere and a non-spherical metaball in Sec.~\ref{sec:validation}. The significance of capturing particle shapes is demonstrated in Sec.~\ref{sec:examples} by interactions of multiple non-spherical particles. Finally Sec.~\ref{sec:conclusion} presents conclusions for the present work.

\section{Numerical model}\label{sec:mathod}

\subsection{Metaball Discrete Element Method}\label{sec:dem}
\subsubsection{Discrete Element Method}
DEM is a method that solves the individual particle (element) motions directly~\cite{luding2008introduction,solov2017mbn}. The translational motions are described by Newton's equation and rotations are governed by the angular momentum conservation equation:
\begin{equation}
\begin{cases}
m_i\bm{a}_i = m_i\bm{g} + \sum \limits_{j=0}^{N-1} \bm{F}_{ij}^c + \bm{F}_i^h,\\[2ex]
\frac{d}{dt}\left(\textbf{I}_i \bm{\omega}_i\right) = \sum \limits_{j=0}^{N-1} \bm{T}_{ij}^c + \bm{T}_i^h,
\end{cases}
\label{eq:DEM}
\end{equation}
where $N$ is total number of particles, $m_i$ and $\bm{a}_i$ are the mass and acceleration of particle $i$, respectively. Forces acting on particle $i$ include gravitational force $m_i\bm{g}$, the hydrodynamic force $\bm{F}_i^h$ and the contact force $\bm{F}_{ij}^c$ between particle $i$ and $j$. $\textbf{I}_i$ is the inertia tensor and $\bm{\omega}_i$ is the angular velocity. $\bm{T}_i^h$ and $\bm{T}_{ij}^c$ represent torques due to the hydrodynamic and contact forces. It is worth to mention that the angular momentum conservation equation in Eq.~\ref{eq:DEM} is handled by solving Euler's equation~\cite{solov2017mbn} under the body-frame.

Within the DEM formalism, there are various kinds of contact laws to determine the contact force $\bm{F}^c$. One of the most widely used models is the linear spring dashpot model introduced by Cundall and Strack~\cite{cundall1979discrete}. The normal component of $\bm{F}^c$ between particle $i$ and $j$ is given by:
\begin{equation}
F_n^{c} = k_n\delta+\eta_n \left(\bm{v}_j-\bm{v}_i \right)\cdot\bm{n},
\end{equation}
where $\delta$ is the particle overlapping distance, $k_n$ is the normal spring stiffness and $\eta_n$ is the normal damping coefficient. The unit normal vector $\bm{n}$ points from particle $j$ to particle $i$. The tangential contact force $F_t^{c}$ follows Coulomb's law: $F_t^{c} \leq \mu_s F^{cn}$ and can be determined as:
\begin{equation}
\begin{cases}
F_t^{c} &= min(\mu_s F_n^{c}, F_{t0}^{c}),\\[2ex]
F_{t0}^{c} &= \norm{-k_t \bm{\xi} - \eta_t \left(\bm{v}_j-\bm{v}_i \right)\cdot\bm{t}},
\end{cases}
\end{equation}
where $\mu_s$ is the static friction coefficient. $k_t$ and $\eta_t$ are tangential spring stiffness and damping coefficient. The unit tangential vector $\bm{t}$ and tangential spring $\bm{\xi}$ can be determined by relative velocity. More details can be found in~\cite{luding2008introduction,solov2017mbn,chen2020efficient}. The second-order Velocity Verlet scheme is employed to solve Eq.~\ref{eq:DEM} numerically.

\subsubsection{Metaball function}
\begin{figure}[b]
\begin{centering}
\includegraphics{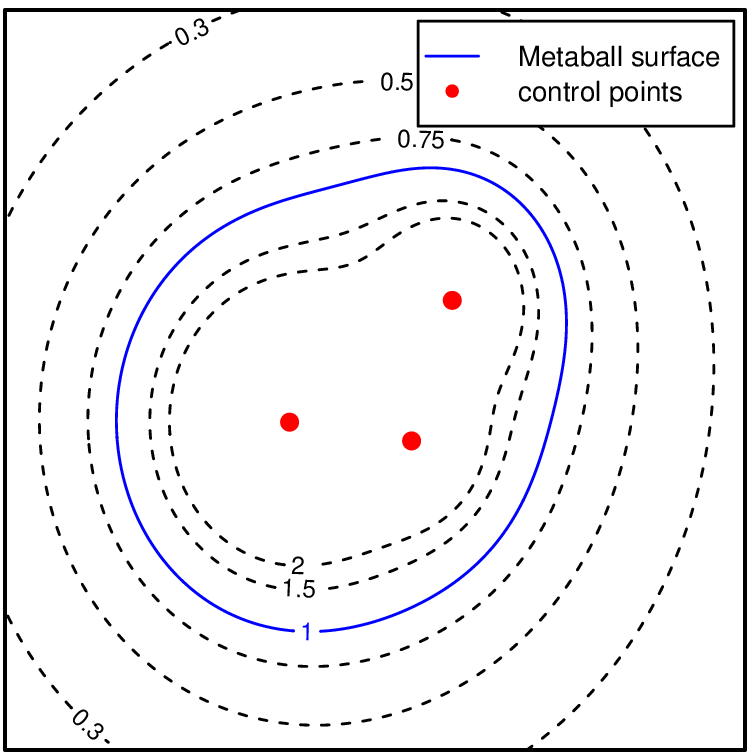}
\caption{2D graphic illustration of a metaball particle and its control points, the dash lines refer to Eq.~\ref{eq:metaball} with different values.}
\label{fig:2d_meta}
\end{centering}
\end{figure}

One key aspect of modern DEM schemes is the shape descriptor for non-spherical particles. Recently, the authors introduced a novel shape descriptor: metaball function~\cite{zhang2021metaball} which can be used for non-spherical particles with round features, such as river pebbles. Metaball function describes particle shapes by an analytical expression, it can be considered as a natural extension of the sphere, ellipsoid etc. The metaball equation used in this study is defined as:

\begin{equation}
M(\bm{x})=\sum_{i=0}^{n-1}{\frac{K_i}{\Vert\bm{x}-\bm{x}_i\Vert^2}}=1
\label{eq:metaball}
\end{equation}
where $\bm{x}_i$ is the $i$th control point which determines the skeleton of the shape, $K_i$ is the corresponding weight which controls the influence range of $\bm{x}_i$, $n$ is the total number of control points. It is clear that a sphere can be described by Eq.~\ref{eq:metaball} with a single control point as centre and $\sqrt{K_0}$ as the radius. One advantage of Eq.~\ref{eq:metaball} is: no constraints on choosing control points and the weights can be any non-negative value. Because of this flexibility, the above metaball equation can be used to describe many complex shapes. Although there are no limitations on using metaball for non-convex shapes, only convex shapes are considered here as the first step. A 2D metaball particle and its control points are illustrated in Fig.~\ref{fig:2d_meta}. The contour plot of $M(\bm{x})$ also highlights that its value decreases with increasing area (volume) when $M(\bm{x})<1$. This property will be used to handle collisions between metaballs later.

\subsubsection{Collision between two metaballs}
The collision algorithm of two metaballs requires defining collision properties including contact point, contact normal and overlaps. To avoid intersections between two metaballs, a sphero-metaball approach is developed in~\cite{zhang2021metaball}, where the original metaball is eroded to an internal metaball with a similar shape and then dilated by a sphere with radius $R_s$. The task of finding the closest points of internal metaballs can be solved as an optimization problem with the help of the analytical expression of metaballs (Eq.~\ref{eq:metaball}). The optimization problem is defined as follows:

\begin{equation}
\begin{aligned}
&Minimize \quad M_0(\bm{x})+M_1(\bm{x}) \\
&subject \text{ } to \quad  c_{tol}<\abs{M_0(\bm{x})}<1, \quad c_{tol}<\abs{M_1(\bm{x})}<1
\end{aligned}
\label{eq:opt}
\end{equation}

where the $M_0(\bm{x})$ and $M_1(\bm{x})$ are the function of two metaballs, $c_{tol}$ is a small tolerance to avoid the solution of $M_0(\bm{x})+M_1(\bm{x})=0$ when $\norm{\bm{x}}\to \infty$. If the solution of local minimum exists, the gradient of Eq.~\ref{eq:opt} must equal to zero:
\begin{equation}
\nabla (M_0(\bm{x})+M_1(\bm{x}))=\bm{0}
\label{eq:grad}
\end{equation} 

Eq.~\ref{eq:grad} is solved by the Newton-Raphson method numerically in this study, more details can be found in~\cite{zhang2021metaball}. Once the local minimum point $\bm{x}_m$ is found, the closest points on metaballs ($\bm{x}_{c0}$ and $\bm{x}_{c1}$) are approximated by the intersection points between the line through $\bm{x}_m$ with direction $\nabla M(\bm{x}_m)$ as shown in Eq.~\ref{eq:cp}.

\begin{equation}
\begin{cases}
& \bm{x}_{c0} = \bm{x}_m + q_0\nabla M_0(\bm{x}_m) \\
& \bm{x}_{c1} = \bm{x}_m + q_1\nabla M_1(\bm{x}_m)
\end{cases}
\label{eq:cp}
\end{equation}
By using Eq.~\ref{eq:cp} and Taylor series expansion of $M_0(\bm{x}_{c0})$ about point $\bm{x}_m$, ignoring second and higher order terms, and combined with $M_0(\bm{x}_{c0})=1$, $q_0$ can be expressed explicitly as:

\begin{equation}
q_0 = \frac{1-M_0(\bm{x}_m)}{\norm{\nabla M_0(\bm{x}_m)}^2}
\label{eq:k0}
\end{equation}
$q_1$ can be calculated in the same way. Eq.~\ref{eq:cp} and \ref{eq:k0} are fairly accurate when particles are close to contact and tend to overestimate minimum distance when particles are far away since $M(\bm{x})$ quadratically decreases with distance. In another word, no collision happens when the error of Eq.~\ref{eq:cp} is large. Finally, the overlap $\delta$, contact direction $\bm{n}$ and contact point $\bm{x}_{cp}$ are defined as:

\begin{equation}
\begin{cases}
& \delta = R_{s0}+R_{s1}-\norm{\bm{x}_{c1} - \bm{x}_{c0}} \\
& \bm{n} = \frac{\bm{x}_{c0} - \bm{x}_{c1}}{\norm{\bm{x}_{c0} - \bm{x}_{c1}}} \\
& \bm{x}_{cp} = \bm{x}_{c0} + (R_{s0}-0.5\delta)\bm{n}
\end{cases}
\label{eq:contactprop}
\end{equation}

\begin{figure}[t]
\begin{centering}
\includegraphics[width=0.6\linewidth]{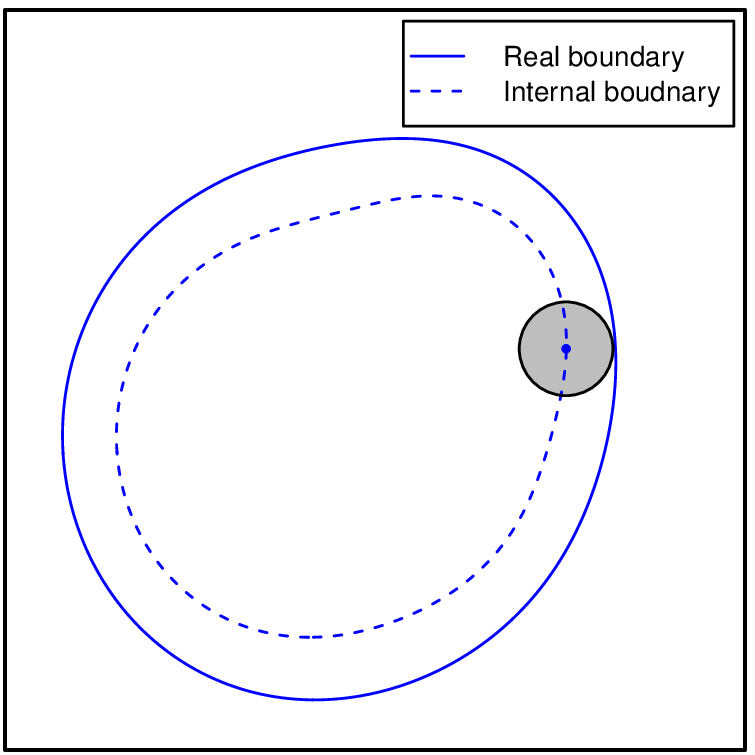}
\caption{An illustration of a sphero-metaball, the solid and dash lines refer to original metaball and internal metaball. The original metaball is approximated by the Minkowski sum of the internal metaball and a sphere.}
\label{fig:sphere_meta}
\end{centering}
\end{figure}

\subsubsection{Collision between metaball and plane}
\begin{figure}[t]
\begin{centering}
\includegraphics[width=0.6\linewidth]{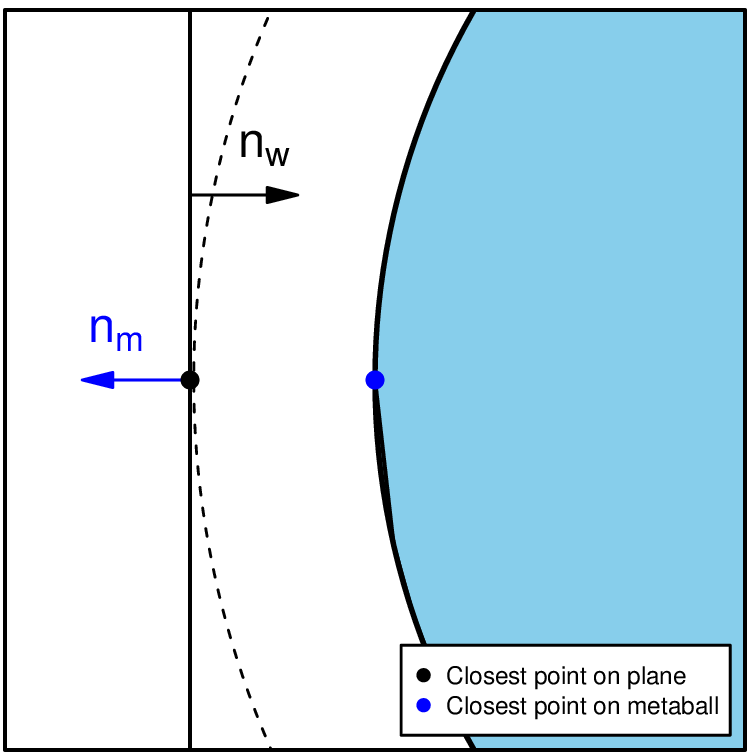}
\caption{Illustration of the contact scheme for metaball-plane collision.}
\label{fig:metawall}
\end{centering}
\end{figure}

The collisions between metaball and plane are handled similarly to metaball collisions. The task becomes: finding the closest points between metaball and plane. It is equal to finding a point on metaball with a normal that is perpendicular to the plane and pointing towards the inside of the plane. This problem can be simplified by rotating the coordinate around an internal point (usually the mass centre) of the metaball to make sure that the norm of the plane is perpendicular to the x-axis (Fig.~\ref{fig:metawall}). Since the distance between metaball and plane is small, the problem can be further modified as finding a point $\bm{x}_{cw}$ on the plane where its normalized gradient regarding the rotated metaball function $M^{R}$ equals to $(-1,0,0)$. Thus we have:
\begin{equation}
\begin{cases}
& \frac{\partial}{\partial y} M^{R}(\bm{x}^{R}_{cw}) = 0\\
& \frac{\partial}{\partial z} M^{R}(\bm{x}^{R}_{cw}) = 0\\
& \frac{\partial}{\partial x} M^{R}(\bm{x}^{R}_{cw}) < 0
\end{cases}
\label{eq:metawall}
\end{equation}
Eq.~\ref{eq:metawall} is also solved by the Newton-Raphson Method. The previous $\bm{x}^{R}_{cw}$ is used as the initial point if this potential collision already exists at the previous time step. Otherwise, the initial point is determined by projecting the control point with the smallest distance to the plane.

Once $\bm{x}^{R}_{cw}$ is determined, the corresponding closest point on metaball $\bm{x}^{R}_{cm}$ is found as: 

\begin{equation}
\begin{aligned}
\bm{x}^{R}_{cm} & = \bm{x}^{R}_{cw} + q^R\nabla M^{R}(\bm{x}^{R}_{cw})
\end{aligned}
\label{eq:metawallcp}
\end{equation}
using $M^{R}(\bm{x}^{R}_{cm})=1$ and Taylor series expansion, we have:

\begin{equation}
q^R = \frac{1-M^R(\bm{x}^{R}_{cw})}{\norm{\nabla M^R(\bm{x}^{R}_{cw})}^2}
\label{eq:k}
\end{equation}
The closest points are then rotated back to the global coordinate and the overlap $\delta$, contact direction $\bm{n}$ and contact point $\bm{x}_{cp}$ are defined as:

\begin{equation}
\begin{cases}
& \delta = R_{s}-\norm{\bm{x}_{cm} - \bm{x}_{cw}} \\
& \bm{n} = \frac{\bm{x}_{cm} - \bm{x}_{cw}}{\norm{\bm{x}_{cm} - \bm{x}_{cw}}} \\
& \bm{x}_{cp} = \bm{x}_{cw} + 0.5\delta\bm{n}
\end{cases}
\label{eq:contactprop2}
\end{equation}
It is worth mentioning that this collision algorithm could potentially be used to couple the Metaball DEM with traditional polyhedral DEM to enhance the modelling capability of this method.

\subsection{Lattice Boltzmann Method}\label{sec:lbm}
\begin{figure}[b]
  \begin{centering}
   \includegraphics[width=0.6\linewidth]{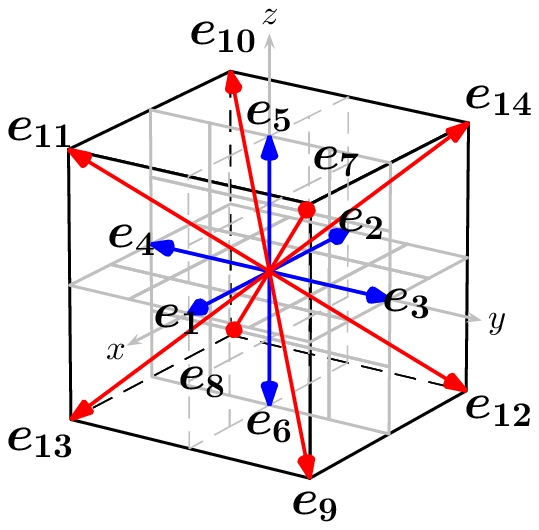}
   \caption{Discrete velocity vectors for D3Q15~\cite{galindo2013coupled}.}
   \label{fig:lbmcell}
  \end{centering}
\end{figure}

The fluid flow is simulated by the Lattice Boltzmann equation (LBE) – a discretized form of the Boltzmann equation~\cite{galindo2012numerical,galindo2013coupled, galindo2013lattice} and the D3Q15 lattice model is used, where the space is divided into cubic lattices. The velocity domain is discretized to fifteen velocity vectors as shown in Figure~\ref{fig:lbmcell}. The discrete velocity vectors are defined as follows:
\begin{equation*}
\resizebox{0.8\hsize}{!}
{$\bm{e}_i = \left\{ 
  \begin{array}{l l l}
    0, & \quad \text{$i$ = 0,}\\[0.5ex]
    (\pm C,0,0),(0,\pm C,0),(0,0,\pm C), & \quad \text{$i$ = 1 to 6,}\\[0.5ex]
    (\pm C,\pm C,\pm C), & \quad \text{$i$ = 7 to 14,}\\
  \end{array} \right.$}
\end{equation*}
where $C=\Delta{x}_{LBM}/\Delta{t}_{LBM}$ being the characteristic lattice velocity, $\Delta{x}_{LBM}$ and $\Delta{t}_{LBM}$ are the lattice size and time step of LBM.

Based on the Chapman-Enskog expansion of the LBE, an evolution rule is applied to every distribution function~\cite{guo2002discrete}:
\begin{equation}
f_i(\bm{x}+\bm{e}_i\Delta{t}_{LBM},t+\Delta{t}_{LBM}) = f_i(\bm{x},t) + \Omega_{col},
\end{equation}
where $f_i$ is the probability distribution function, $\bm{x}$ is the position of the local lattice, $\Omega_{col}$ is the collision operator. The well-known Bhatnagar-Gross-Krook (BGK) collision operator is used in this study,
\begin{equation}
\Omega_{col} = \frac{\Delta{t}_{LBM}}{\tau}(f^{eq}_{i}-f_i),
\label{eq:lbm_collide}
\end{equation}
where $\tau$ is the relaxation time and $f^{eq}_{i}$ is the equilibrium distribution given by
\begin{equation}
f^{eq}_{i} = \omega_i\rho_f \bigg(1 + 3\frac{\bm{e}_i \cdot \bm{u}_f}{C^2} + \frac{9(\bm{e}_i \cdot \bm{u}_f)^2}{2C^4} - \frac{3u_f^2}{2C^2}\bigg),
\end{equation}
The weights are $\omega_0=2/9$, $\omega_i=1/9$ for $i=$1 to 6, $\omega_i=1/72$ for $i=$7 to 14. The kinetic viscosity is related to the relaxation time by
\begin{equation}
\nu = \frac{\left(\Delta x_{LBM}\right)^2}{3\Delta t_{LBM}}\bigg(\tau - \frac{1}{2}\bigg),
\end{equation}
here the Mach number is defined as the ratio of the maximum fluid velocity to $C$. When $Ma\ll1$, the LBE can be recovered to the Navier-Stokes equation. More detail can be found in~\cite{mohamad2011lattice}. The macroscopic properties of fluid such as density $\rho_f$ and flow velocity $\bm{u}_f$ can be determined by the zeroth and the first-order moment of the distribution function: 
\begin{equation}
\begin{array}{ll}
\rho_f(\bm{x}) &= \sum_{i=0}^{14} f_i(\bm{x}),\\[2ex]
\bm{u}_f(\bm{x}) &= \frac{1}{\rho_f(\bm{x})} \sum_{i=0}^{14} f_i(\bm{x})\bm{e}_i .
\end{array}
\label{eq:blm_rhov}
\end{equation}

\subsection{Coupling scheme between MDEM and LBM}
To successfully model fluid-structure interactions, the no-penetration non-slip boundary conditions need to be imposed on the fluid-solid interface, and the hydrodynamic forces acting on particles are also required.

\subsubsection{Interpolated Bounce Back scheme for Moving boundary condition}
\begin{figure}[t]
\begin{centering}
\includegraphics[width=0.8\linewidth]{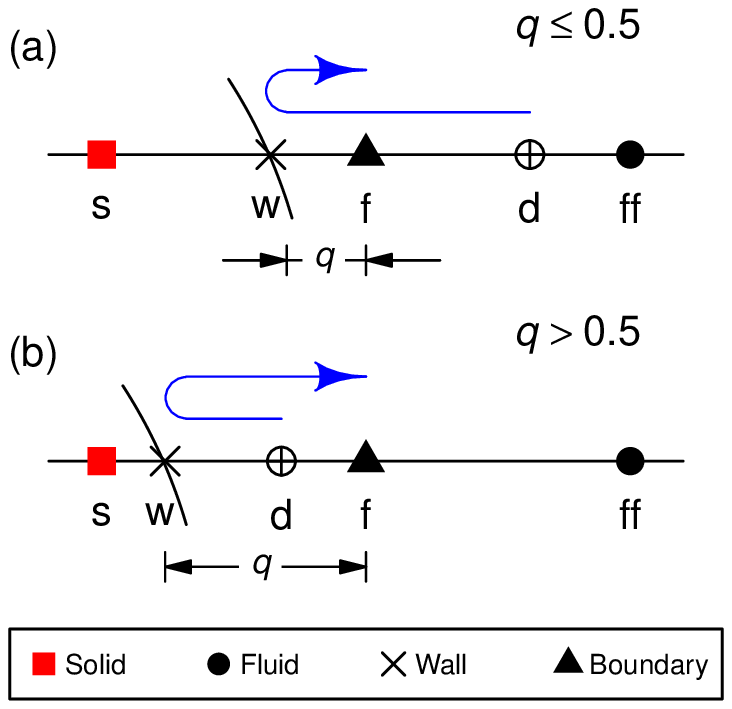}
\caption{Schematic of the interpolated bounce back role at the fluid-structure interface, where ``s" for the closest solid node, ``w" for wall, ``f" for the boundary node, ``ff" for the neighbouring fluid node of ``f".}
\label{fig:ibb}
\end{centering}
\end{figure}

The LBM nodes are divided into fluid nodes and solid nodes, the fluid nodes which are next to the solid boundary are further identified as boundary nodes ($f$ in Fig.~\ref{fig:ibb}). Since the uniform-sized mesh is used in classic LBM, the curved boundaries are generally located between boundary nodes and solid nodes. Thus, the distribution functions at boundary nodes that streamed from solid nodes are missing, the critical task is to determine the missing distribution functions properly. 

The simplest solution is the bounce-back role where molecules depart from $\bm{x}_f$ with velocity $\bm{e}_{i'}$ hit on wall and return back to $\bm{x}_f$ with opposite discrete velocity $\bm{e}_{i}$. It is clear that the wall is assumed to be located at the middle point between $\bm{x}_f$ and $\bm{x}_s$ regardless of the actual position, where $\bm{x}_s$ is the neighbour solid node. This assumption leads to stair-wise boundaries which damage the second-order accuracy of LBM. Therefore, interpolated bounce-back(IBB) schemes~\cite{bouzidi2001momentum} are proposed to reduce geometrical errors. The idea is to interpolate the missing distribution functions from existing ones and the interpolation weights depend on the distance $q=\Vert\bm{x}_f-\bm{x}_w\Vert/\Vert\bm{x}_f-\bm{x}_s\Vert$, where $\bm{x}_w$ is the intersection point between the solid surface and discrete velocity. The original IBB scheme needs to treat $q \leqslant 0.5$ and $q>0.5$ conditions separately, Yu et al.~\cite{yu2003viscous} proposed an unified IBB scheme where the distributions at solid boundary $f_{i'}(\bm{x}_w,t+\Delta{t}_{LBM})$ are evaluated first, then the bounce-back role is applied, the missing distributions at $\bm{x}_f$ after streaming $f_{i}(\bm{x}_f,t+\Delta{t}_{LBM})$ is interpolated between $f_{i}(\bm{x}_w,t+\Delta{t}_{LBM})$ and $f_{i}(\bm{x}_{ff},t+\Delta{t}_{LBM})$. 

However, it is found that classical IBB schemes cannot guarantee non-slip conditions at solid surfaces, particularly, at high Reynolds number. Recently, a velocity interpolation-based bounce back scheme (VIBB) is proposed to to reduce the slipping error~\cite{zhang2019velocity}. VIBB scheme is based on the following observation: we can always find a point $\bm{x}_d$ where the distribution departs from $\bm{x}_d$ will arrive at $\bm{x}_f$ after stream and bounce back. The unknown $f_{i}(\bm{x}_f,t+\Delta{t}_{LBM})$ is determined as:
\begin{equation}
f_{i}(\bm{x}_f,t+\Delta{t}_{LBM}) = f^{+}_{i'}(\bm{x}_d,t) + 6\omega_{i'}\rho_{f}\frac{\bm{e}_i \cdot \bm{u}_{w}}{C^2},
\label{eq:eqneq}
\end{equation}
the particle surface velocity is given as: $\bm{u}_{w} = \bm{v}_{pj} + \bm{w}_{pj} \times (\bm{x}_w-\bm{x}_{pj})$, where $\bm{v}_{pj}$ and $\bm{w}_{pj}$ are the translational and angular velocity at the $j$th particle's centroid $\bm{x}_{pj}$, respectively. $f^{+}_{i'}(\bm{x}_d,t)$ is decomposed into equilibrium $f^{eq}$ and non-equilibrium part $f^{neq}$:
\begin{equation}
f^{+}_{i'}(\bm{x}_d,t) = f^{eq}_{i'}(\rho_d,\bm{u}_d) + f^{neq}_{i'}(\bm{x}_d,t),
\label{eq:eqneq}
\end{equation}
notice that the distributions are dominated by the equilibrium part since the variations of $f^{neq}$ are one order smaller than $f^{eq}$. Thus, it is safe to interpolate/extrapolate $f^{neq}$ and $\rho$ with second-order accuracy~\cite{guo2002extrapolation}:
\begin{equation}
f^{neq}_{i'}(\bm{x}_d,t) = 2q(f^{+}_{i'}\left(\bm{x}_f,t)-f^{eq}_{i'}(\bm{x}_f,t)\right) + (1-2q)(f^{+}_{i'}(\bm{x}_{ff},t)-f^{eq}_{i'}(\bm{x}_{ff},t)),
\label{eq:neqd}
\end{equation}
the density at $\bm{x}_d$ is evaluated as:
\begin{equation}
\resizebox{0.6\hsize}{!}
{$\rho_d = \left\{ 
  \begin{array}{l l}
    \text{$2q\rho_f + (1-2q)\rho_{ff}$}, & \quad \text{$q \leqslant 0.5$,}\\[3ex]
     \text{$\rho_f$}, & \quad \text{$q > 0.5$.}\\[0.ex]
  \end{array} \right.$}
\label{eq:rhod}
\end{equation}
Based on the above analysis, $\bm{u}_d$ play the most important roles in determining unknown distributions. Fortunately, both $\bm{u}_w$, $\bm{u}_f$ and $\bm{u}_{ff}$ are known. $\bm{u}_d$ in Eq.~\ref{eq:eqneq} can be evaluated by linear interpolation separately:
\begin{equation}
\resizebox{0.6\hsize}{!}
{$\bm{u}^{*}_d = \left\{ 
  \begin{array}{l l}
    \text{$2q\bm{u}_f + (1-2q)\bm{u}_{ff}$}, & \quad \text{$q \leqslant 0.5$,}\\[3ex]
     \text{$\frac{1-q}{q}\bm{u}_f + \frac{2q-1}{q}\bm{u}_w$}, & \quad \text{$q > 0.5$,}\\[0.ex]
  \end{array} \right.$}
\end{equation}
or linearly interpolated between $\bm{u}_w$ and $\bm{u}_{ff}$ regardless of $\bm{u}_{f}$:
\begin{equation}
\bm{u}^{**}_d = \frac{1-q}{1+q}\bm{u}_{ff} + \frac{2q}{1+q}\bm{u}_w,
\label{eq:neqd}
\end{equation}
here, we determine $\bm{u}_d$ by weighted averaging $\bm{u}^{*}_d$ and $\bm{u}^{**}_d$:
\begin{equation}
\bm{u}_d = \frac{1}{3}\bm{u}^{*}_d + \frac{2}{3}\bm{u}^{**}_d.
\label{eq:neqd}
\end{equation}

\begin{figure}
\centering     %%% not \center
\subfigure{\label{fig:a}\includegraphics{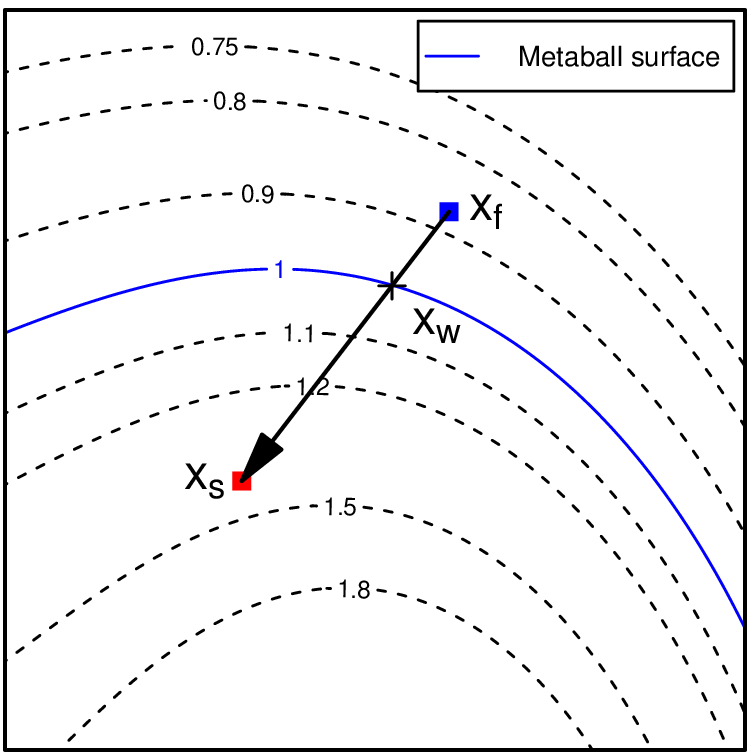}}
\caption{The illustration of intersection between LBM discrete velocity (black arrow) and metaball (blue curve). The contour plot shows that the mataball function value can be used to calculate the intersection points.}
\label{fig:meta_q}
\end{figure}

To couple MDEM with LBM, the parameter $q$ needs to be determined. The intersection point $\bm{x}_{w}$ between metaballs $M(\bm{x})$ and LBM discrete velocities $\bm{e}_i$ must satisfy: $M(\bm{x}_{w}) = c_0$, where $c_0$ is a special metaball function value that depends on spherical radius $R_s$. In practice, $c_0$ is determined by the minimum function value for the  particle surface (see Fig.~\ref{fig:sphere_meta}). $q$ can be calculated by solving following equation:
\begin{equation}
M(\bm{x}_{f}+q \bm{e}_i) = c_0,
\label{eq:calq}
\end{equation}
unfortunately, the solution of Eq.~\ref{eq:calq} does not have an explicit form in general. Although iterative methods such as the Newton-Raphson method can be used, the high computational costs make them less favourable since $q$ has to be updated for every time step. Here, we propose a simple approximation for $q$:
\begin{equation}
q = \frac{c_0-M(\bm{x}_{f})}{M(\bm{x}_{s})-M(\bm{x}_{f})}.
\label{eq:appq}
\end{equation}
As illustrated in Fig.~\ref{fig:meta_q}, if the LBM lattice size is considerably smaller than particle size, it is reasonable to assume that the metaball function decreases linearly with increasing solid surface distance. Eq.~\ref{eq:appq} also guarantees that $q\in [0,1]$. The accuracy of Eq.~\ref{eq:appq} is examined by comparing with Newton's method (with tolerance $10^{-7}$ and maximum iteration number 100), it is found the relative error is around 5\%. The settling velocity from both methods are identical, implying that Eq.~\ref{eq:appq} is a good approximation to determine $q$.

\subsubsection{Momentum Exchange Method for hydrodynamic forces}
The influence of solid particles on the fluid is modelled by the above no-penetration non-slip boundary conditions, particles interact with fluid through the hydrodynamic force $\bm{F}^h$ and torque $\bm{T}^h$ which appear in Eq.~\ref{eq:DEM}. Accurate and efficient calculations of $\bm{F}^h$ and $\bm{T}^h$ are essential for a successful coupling scheme. One widely used scheme is the momentum exchange method (MEM)~\cite{ladd1994numerical}, where the hydrodynamic forces can be calculated as a sum of all the momentum exchanges along with every discrete velocity that collides with solid surfaces. MEM is extensively used for fluid-particle interactions. However, it suffers from numerical noises which introduce extreme flocculating hydrodynamic forces~\cite{peng2016implementation}. Wen et al.~\cite{wen2014galilean} showed that the original momentum exchange method does not obey the Galilean invariance principle. They further proposed a Galilean invariant momentum exchange method which relief numerical noises considerably. Therefore, the Galilean invariant momentum exchange method is adapted in this work, the hydrodynamic force and torque that act on the $j$th particle are given as:
\begin{equation}
\bm{F}^{h}_j = \sum_{i \in \Gamma_j} \left[ (\bm{e}_i-\bm{u}_w)f_i(\bm{x}_f, t) - (\bm{e}_{i'}-\bm{u}_w)f_{i'}(\bm{x}_f,t) \right],
\label{eq:mem}
\end{equation}
\begin{equation}
\bm{T}^{h}_j = \sum_{i \in \Gamma_j} \left( \bm{x}_w - \bm{x}_{pj} \right) \times \left[ (\bm{e}_i-\bm{u}_w)f_i(\bm{x}_f, t) - (\bm{e}_{i'}-\bm{u}_w)f_{i'}(\bm{x}_f,t) \right],
\label{eq:mem}
\end{equation}
where $\Gamma_j$ represents the set of all the discrete velocities that intersect with the $j$th particle. Compared with the original momentum exchange method, $\bm{e}_i$ is shifted by the solid velocity.

\subsubsection{Local refilling algorithm for new fluid nodes}
One drawback of having sharp solid boundaries is that solid nodes may switch to fluid nodes with no fluid information since particles can freely move within the fluid domain. Therefore, these new fluid nodes need to be initialized with proper distribution functions. This procedure is often referred to refilling algorithm. Peng et al.~\cite{peng2016implementation} discussed the influence of different refilling algorithms in terms of numerical stability and accuracy, their results showed that refilling may have significant contributions to the numerical noise on the flocculating hydrodynamic forces. Most refilling algorithms require interpolating/extrapolating information from neighbour nodes. Here, a local refilling algorithm is proposed. The proposed algorithm is based on the following observation: the new fluid node is always close to the solid surface due to the low Mach number requirement of LBM. Therefore, it is reasonable to apply bounce back role for these missing distribution functions that their opposite distribution is known after streaming:
\begin{equation}
f_{i}(\bm{x}_{new}, t) = f_{i'}(\bm{x}_{new}, t) + 6\omega_{i'}\rho_{f}\frac{\bm{e}_i \cdot \bm{u}_{w}}{C^2},
\end{equation}
where $\bm{x}_{new}$ is the new fluid node position. If $f_{i'}(\bm{x}_{new}, t)$ does not exist, the equilibrium refilling is used: $f_{i}(\bm{x}_{new}, t) = f^{eq}(\rho_0, \bm{u}_{new})$, where $\bm{u}_{new} = \bm{v}_{pj} + \bm{w}_{pj} \times (\bm{x}_{new}-\bm{x}_{pj})$. It is obverse that the proposed refilling algorithm is a local scheme and does not depend on interpolations/extrapolations. After distribution functions are refilled, the macroscopic properties like density and velocity are calculated as Eq.~\ref{eq:blm_rhov}.

\subsubsection{Sub-cycling time integration}
There are two time steps involve in DEM-LBM coupling scheme. The time step of DEM $\Delta t_{DEM}$ is typically around~$10^{-6}$ \si{s} and it is easy to reach~$10^{-7}$ \si{s} to make sure the contact is properly resolved in time. On the other hand, the time step of LBM $\Delta t_{LBM}$ is often found severe orders of magnitude larger than the DEM one since it is a function of the viscosity as in Eq.~\ref{eq:lbm_collide}. Therefore, the sub-cycling time integration proposed by Feng et al.~\cite{feng2007coupled} is used in this study. After one LBM computational step, the hydrodynamic force and torque are assumed unchanged and a sub-cycling is used to update contact forces, particle positions, and velocities. The sub-cycling step is defined as: $n_s=\Delta t_{LBM}/\Delta t_{DEM}$. We found $n_s$ has little influence on the overall accuracy if $\Delta t_{DEM}$ is small enough to guarantee a decent contact resolution.

\subsubsection{Special treatments to handle low resolution between particles}\label{sec:treatment}
IBB and MEM enjoy high accuracy due to the sharp interface representation but suffer from numerical stability issues for the same reason, particularly, when simulations involve multiple particles. Therefore, diffuse interface based schemes like Immersed Boundary Method, Immersed Moving Boundary method are often used for DEM-LBM coupling despite their non-physical interface representation. Here, we show that the numerical stability of IBB and MEM can be significantly enhanced with proper treatments.

1. When a LBM node lays between two particles, there may not have enough fluid nodes to conduct VIBB, then halfway bounce-back is used in this case.

\begin{figure}[t]
\begin{centering}
\includegraphics[width=0.8\linewidth]{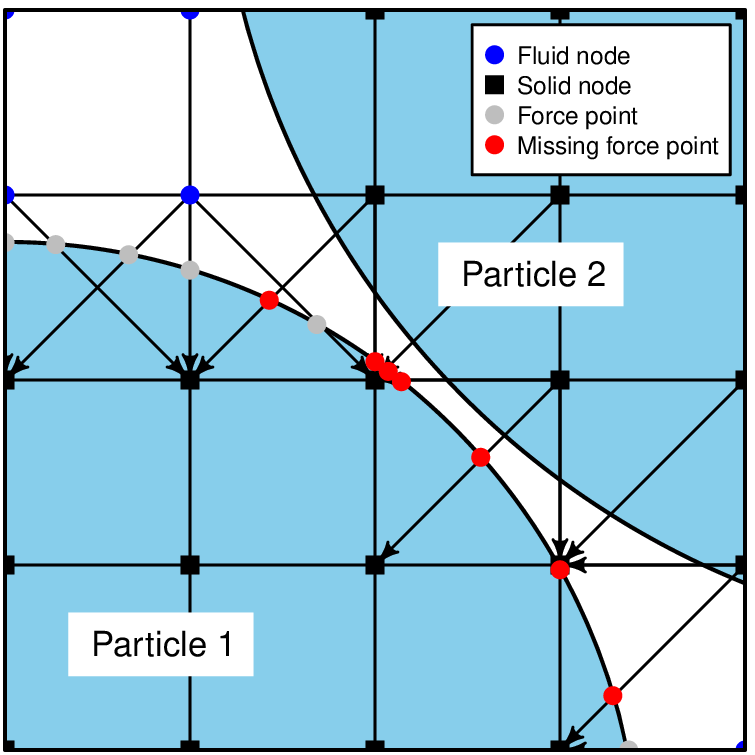}
\caption{When the gap between particles is small, some neighbouring nodes of particles are covered by other particles, caused an anisotropic distribution of force points.}
\label{fig:force_point}
\end{centering}
\end{figure}

2. When the particle-particle gap or particle-wall gap is small, the distribution of force points on the particle surface becomes less isotropic since only the momentum exchange from fluid nodes is considered (see Fig.~\ref{fig:force_point}). The anisotropic effect can introduce a significant disturbance to particle dynamics. We found that the key to restore the isotropic is to take into account the momentum exchange on the missing force points. Due to the limited information of fluid, the distribution function at missing force points is reconstructed as equilibrium distribution with the initial density $\rho_0$ and $\bm{u}_w$. The hydrodynamic force and torque at the missing force points are then evaluated as: 
\begin{equation}
\bm{F}^{h,i}_j = (\bm{e}_i-\bm{u}_w)f^{eq}_i(\rho_0,\bm{u}_w) - (\bm{e}_{i'}-\bm{u}_w)f^{eq}_{i'}(\rho_0,\bm{u}_w),
\label{eq:fh_s_issue}
\end{equation}
\begin{equation}
\bm{T}^{h,i}_j = \left( \bm{x}_w - \bm{x}_{pj} \right) \times \left[ (\bm{e}_i-\bm{u}_w)f^{eq}_i(\rho_0,\bm{u}_w) - (\bm{e}_{i'}-\bm{u}_w)f^{eq}_{i'}(\rho_0,\bm{u}_w) \right].
\label{eq:th_s_issue}
\end{equation}
It is worth to mention that Christoph and Ulrich~\cite{rettinger2022efficient} also report similar issues recently. In their treatment, the missing hydrodynamic force is given by $\bm{F}^{h,i}_j=2w_i \rho_0 \bm{e}_i$ which is consistent with Eq.~\ref{eq:fh_s_issue}. In fact, Eq.~\ref{eq:fh_s_issue} becomes identical to their treatment if $\bm{u}_w=\bm{0}$.

\section{Validation}\label{sec:validation}
\subsection{Settling of a single sphere with metaball equation}\label{sec:settling_sphere}
\begin{figure}[t]
\begin{centering}
\includegraphics[width=0.8\linewidth]{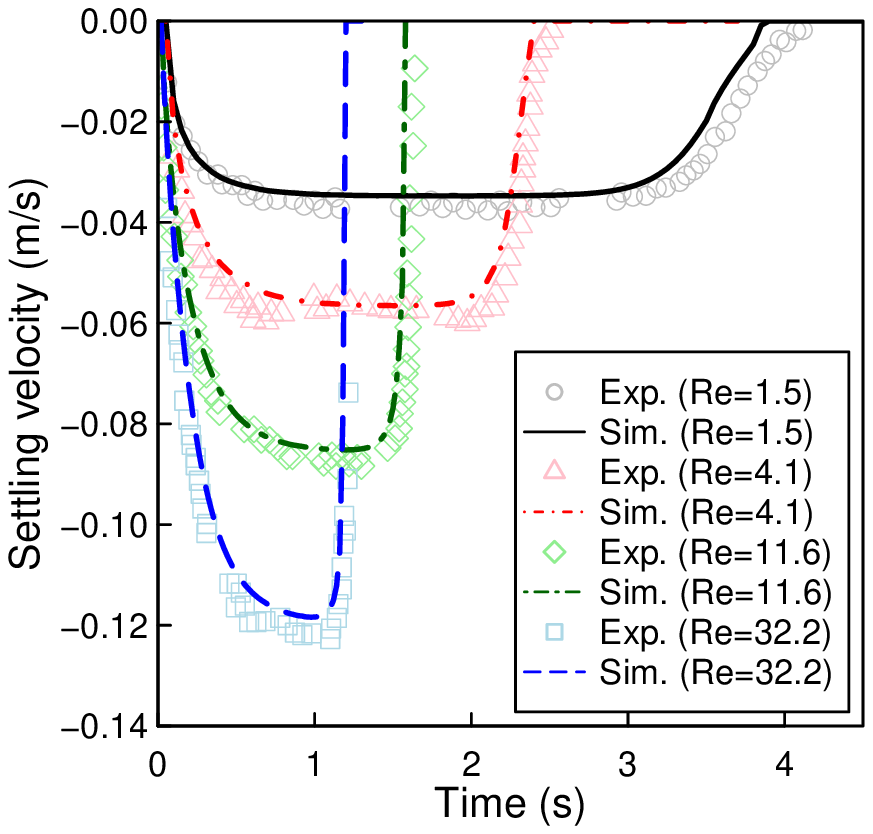}
\caption{Comparison between simulated settling velocities and experimental measurements for a sphere during the settling process.}
\label{fig:settling_vel}
\end{centering}
\end{figure}

To validate the proposed model, the settling of a single sphere in a viscous fluid is simulated to examine the dynamic behaviours of the sphere and associated fluid motion. The time evolution of settling velocities is compared with experimental results. The domain size is $0.1 \times 0.1 \times 0.6$ \si{m}. A sphere particle with diameter $d_p=0.015$ \si{m} and density $\rho_p=1120$ \si{kg/m^3} is placed at a height of $0.12$ \si{m} from the bottom. Four different fluids are used with fluid density $\rho_f= 970, 965, 962, 960$ \si{kg/m^3} and the kinetic viscosity $\nu= 3.85 \times 10^{-4}, 2.2 \times 10^{-4}, 1.17 \times 10^{-4}, 0.6 \times 10^{-4}$ \si{m^2/s} respectively. Non-slip boundary conditions are applied for all boundary walls. Note that in the simulations, the gravitational body force is only applied to the particle, thus, a relative gravity ($(1-\rho_f/\rho_s)g$) is used as suggested by Feng and Michaelides~\cite{feng2004immersed}, where $g=9.8$ \si{m/s^2} is the gravity. The sphere is handled as a metaball instead of using the sphere equation directly. We choose the spherical radius $R_s=1.0 \times 10^{-4}$ \si{m}, the metaball function of a sphere is then given as $M(\bm{x})=k_0/\Vert\bm{x}-\bm{x}_p\Vert^2=1$, where $\bm{x}_p$ is the mass center of the sphere and $k_0=(0.5 d_p-R_s)^2$. Ladd~\cite{ladd1994numerical} suggested that the sphere diameter should be larger than 9 LBM cells to ensure sufficient accuracy. Here, the space step is set as $\Delta x_{LBM}=0.001$ \si{m}, thus, $d_p=15$ under lattice unit. The LBM and DEM time steps are given as: $\Delta t_{LBM}=2.0 \times 10^{-4}$ and $\Delta t_{DEM}=2.0 \times 10^{-6}$ \si{s}. The Reynolds number $Re$ is defined as: $Re=d_p u_t /\nu$, where $u_t$ is the terminal settling velocity. The time series of simulated settling velocities are compared to the experimental data presented in Fig.~\ref{fig:settling_vel}. The good agreements between simulation results and experimental observations suggest that the proposed coupling scheme can accurately capture the fluid-particle interactions.

\subsection{Settling of a non-spherical metaball}\label{sec:settling_metaball}
\begin{figure}[t]
\begin{centering}
\includegraphics[width=0.8\linewidth]{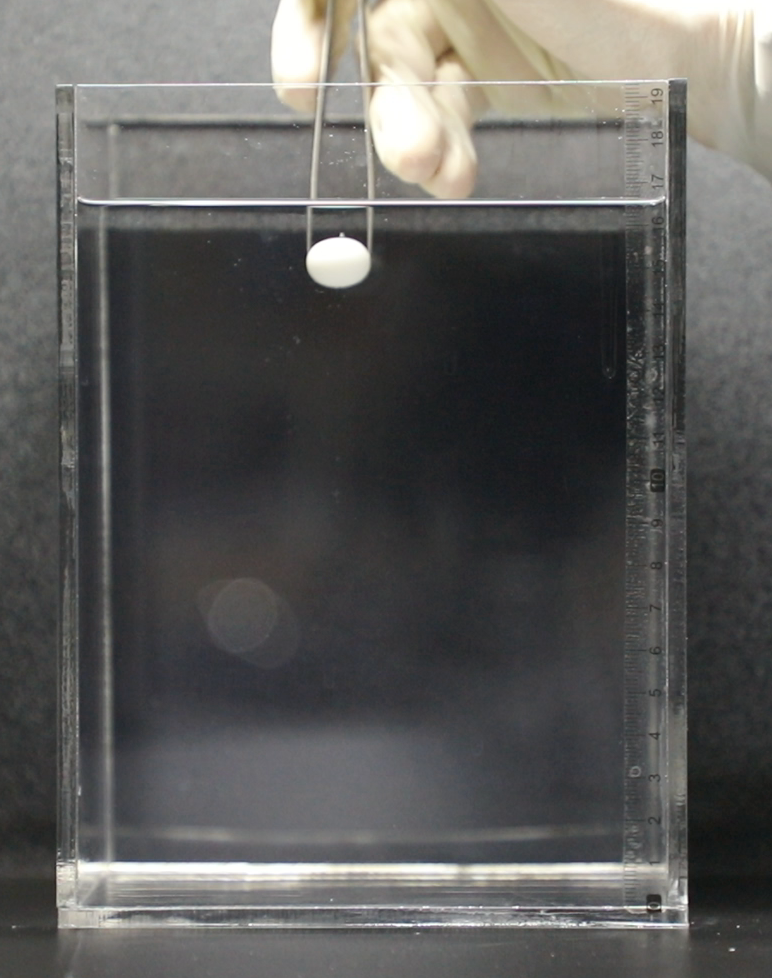}
\caption{Experiment setup for single metaball settling in fluid.}
\label{fig:exp}
\end{centering}
\end{figure}

\begin{figure}
\centering     %%% not \center
\subfigure[Metaball particle in simulation]{\label{fig:a}\includegraphics[width=50mm]{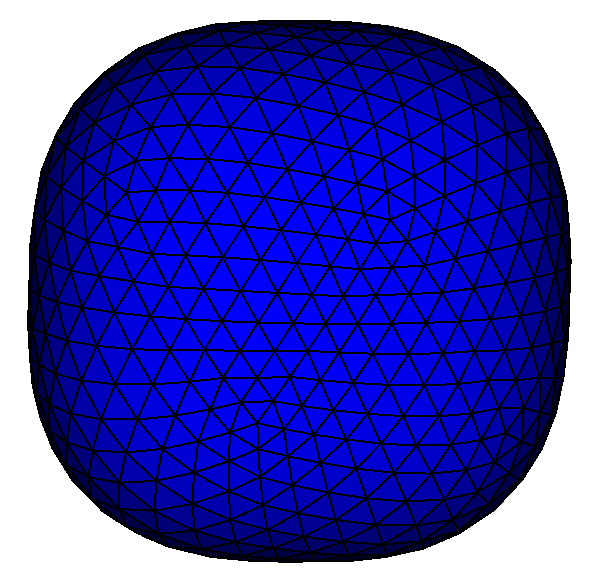}}
\subfigure[Metaball particle in experiment]{\label{fig:a}\includegraphics[width=50mm]{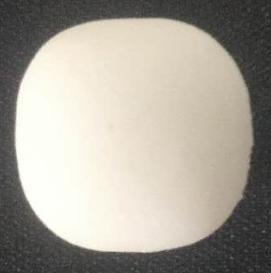}}
\caption{Particle shapes that used in experiments and simulations.}
\label{fig:particle_shape}
\end{figure}

One advantage of metaball DEM is it can be used to describe non-spherical particles with round surfaces. To further validate the proposed MDEM-LBM model, we conducted both experiments and simulations for the settling of a non-spherical metaball in a viscous fluid. The experimental setup is shown in Fig.~\ref{fig:exp}, where a rectangle container is used with dimension of $0.15 \times 0.15 \times 0.2$ \si{m}. The shape of metaball is shown in Fig.~\ref{fig:particle_shape}, its control points form a square and the shape can be quantitatively described by elongation $f_{elong}=0.84375$ which is defined as the ratio of bounding box width and length. The metaball is 3D printed by using a high-resolution surface mesh to conserve geometrical properties and keep a smooth surface. The surface mesh contains 159602 vertices where the metaball descriptor for the same shape only requires 4 control points. It is worth to point out that the surface mesh is only used to visualize the particle shapes and 3D printing, where the proposed collision and coupling algorithm does not require the discretization of particles. Since control points and weights of the metaball are predefined and the metaball is then manufactured by 3D printing, the particle shape in simulations is exact the same as the one in experiments (except for the error from 3D printing). The density and volume of 3D printed particles are $1134.156$ \si{kg/m^3} and $1.648 \times 10^{-6}$ \si{m^3}. The metaball is released by using a pair of tweezers from a completely submerged position and the initial orientation is controlled to make sure that the particle maximum projection area is perpendicular to the settling direction. Two types of Di-methyl silicon oil with viscosity $4.22 \times 10^{-4}$ and $8.91 \times 10^{-5}$ \si{m^2/s} (measured by a LICHEN NDJ-8S viscometer) are used in experiments. The trajectory of the metaball is recorded by a camera (CANON EOS 5D Mark IV with a lens of 24-105mm) with 1080P resolution and 50fps frequency. The videos are post-processed into binary images and the metaball centroid is determined with MATLAB package "regionprops". The metaball settling velocity is calculated by counting pixels between the position change of the centroid between consecutive frames. Same parameters are used in simulations, the space and time step is set as $\Delta x_{LBM}=0.001$ \si{m}, $ \Delta t_{LBM}=2.0 \times 10^{-4}$ and $\Delta t_{DEM}=2.0 \times 10^{-6}$ \si{s}.

\begin{figure}[t]
\begin{centering}
\includegraphics[width=0.8\linewidth]{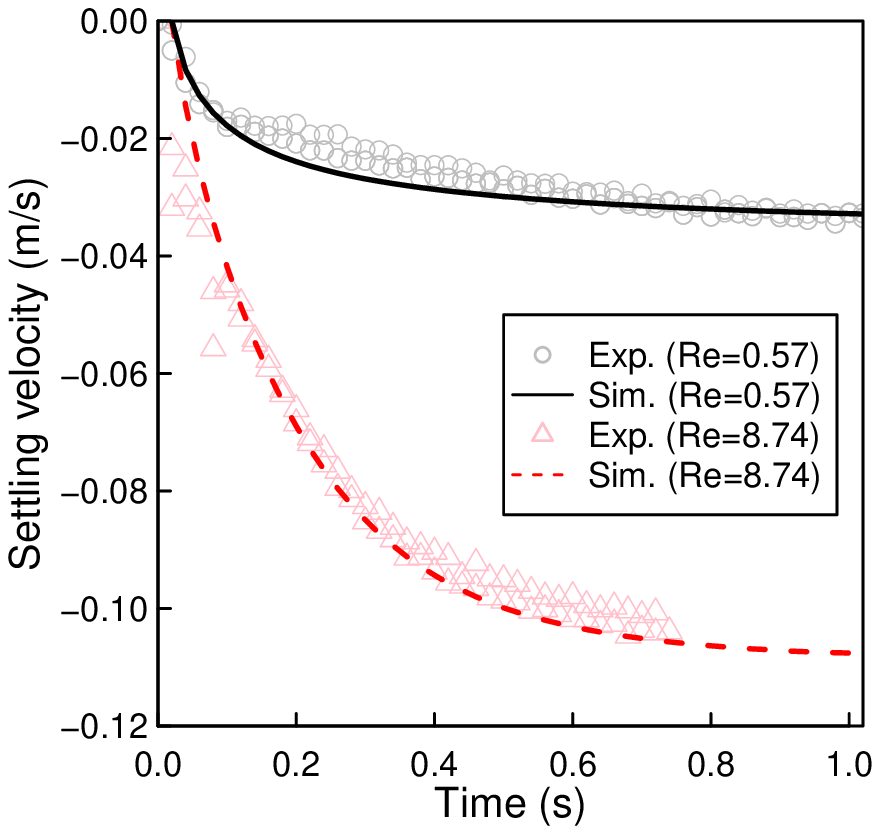}
\caption{Comparison between simulated settling velocities and experimental measurements for a metaball during the settling process.}
\label{fig:meta_settling_vel}
\end{centering}
\end{figure}

\begin{figure}
\centering     %%% not \center
\subfigure{\includegraphics[width=55mm]{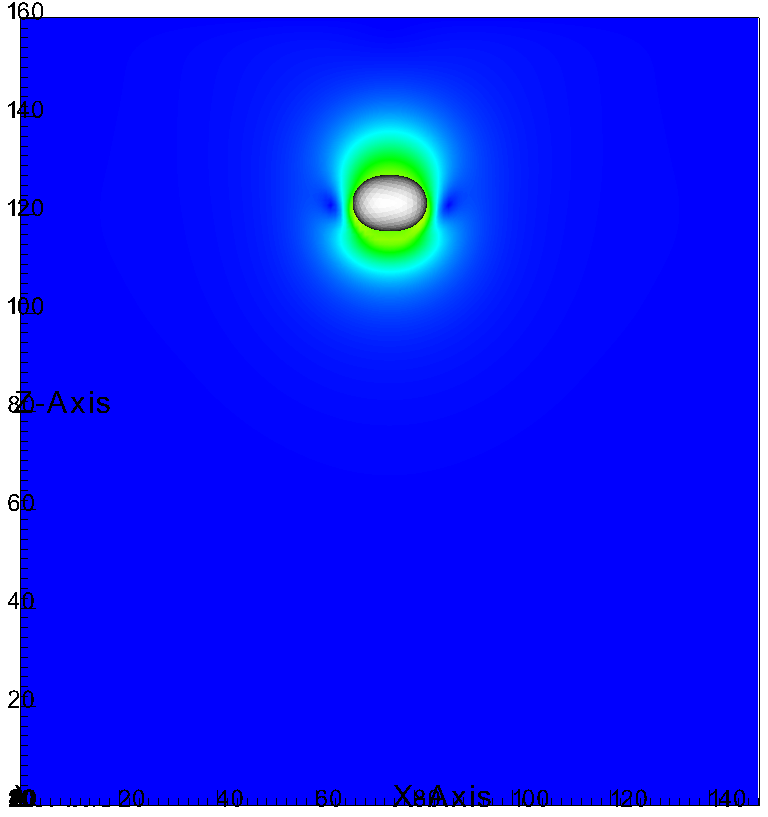}}
\subfigure{\includegraphics[width=55mm]{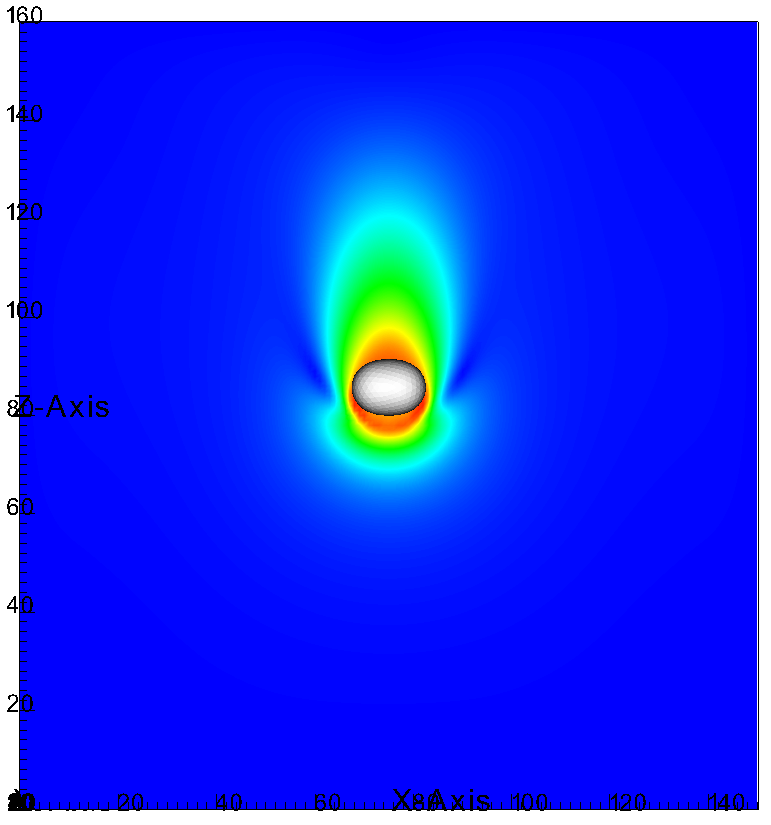}}
\subfigure{\includegraphics[width=55mm]{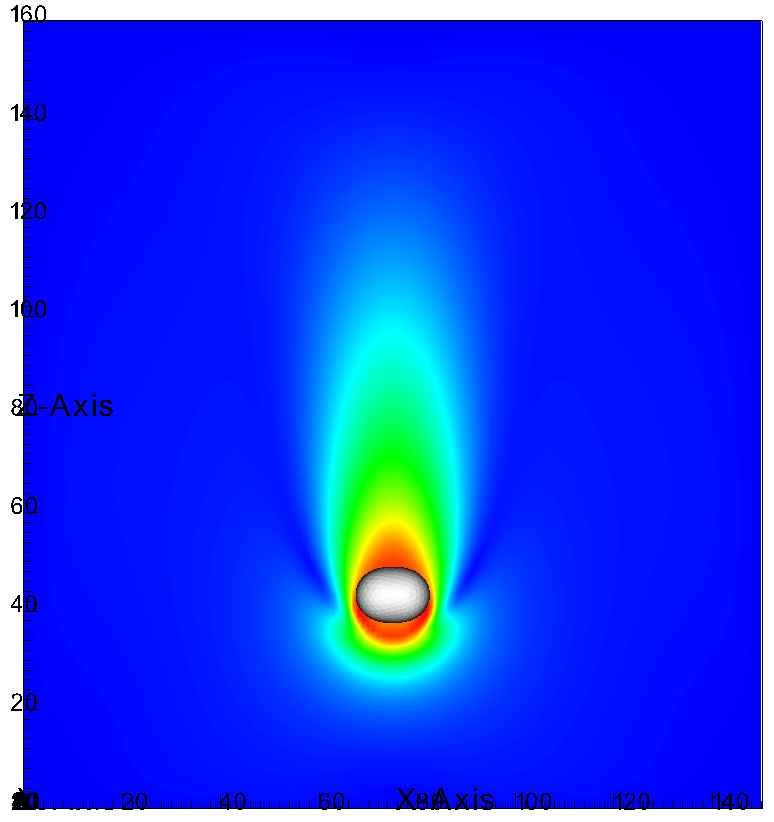}}
\subfigure{\includegraphics[width=55mm]{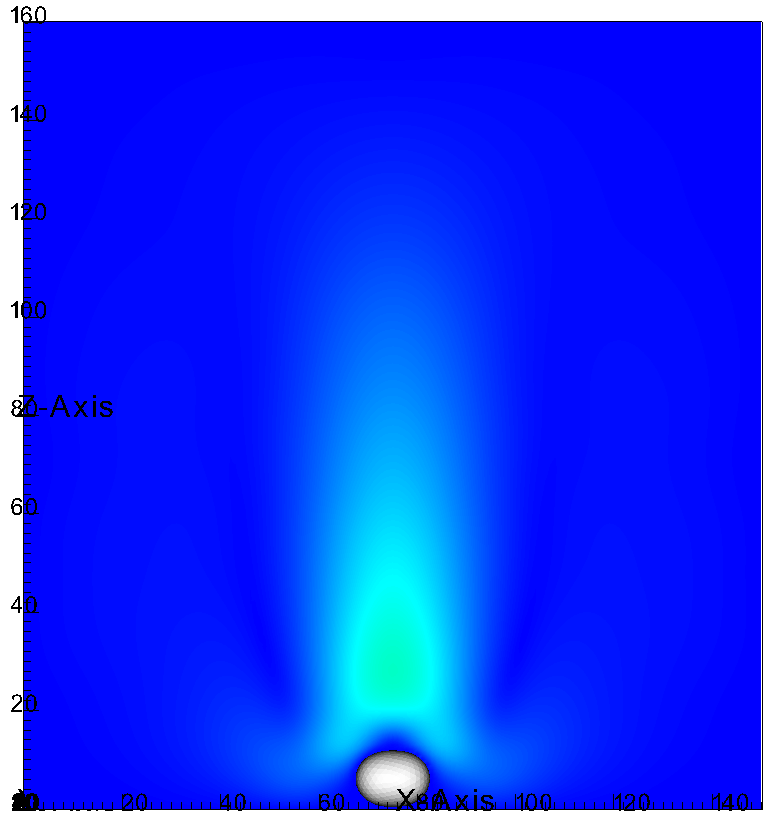}}
\caption{Snapshot of the metaball settling simulation for $Re=8.74$ at 0.2, 0.6, 1 and 1.6 s, colour indicates fluid velocity magnitude.}
\label{fig:a_meta_snapshot}
\end{figure}

\begin{figure}[t]
\begin{centering}
\includegraphics[width=0.8\linewidth]{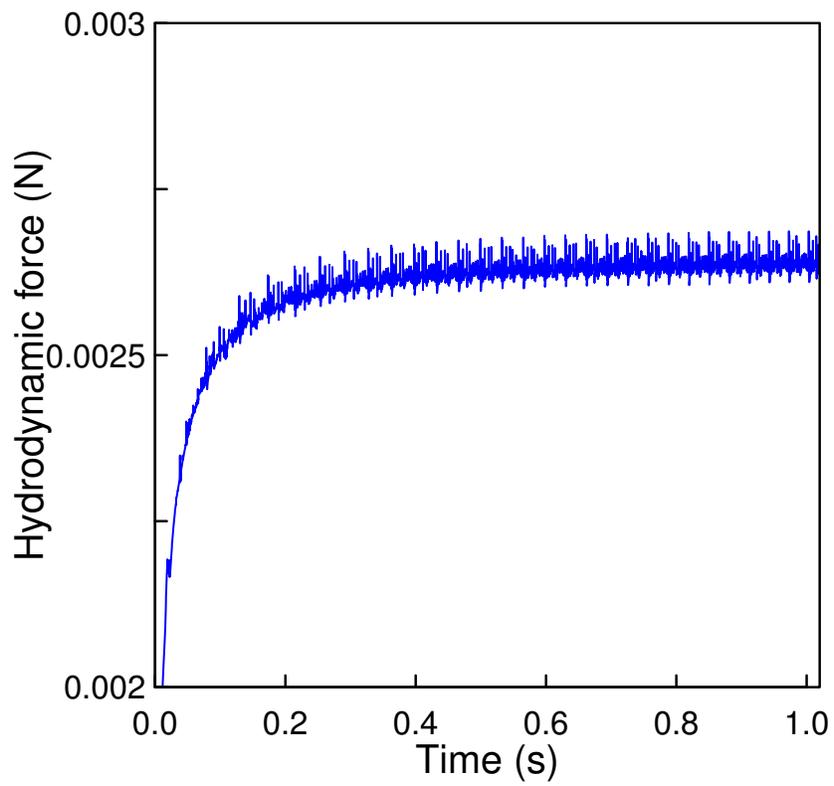}
\caption{Time evolution of the hydrodynamic force acted on a metaball during settling, $Re$=8.74.}
\label{fig:meta_fh}
\end{centering}
\end{figure}

Fig.~\ref{fig:meta_settling_vel} shows the time series of settling velocity for both simulations and experiments. Since $Re$ is relatively low, no rotations are observed if the particle is released with maximum projection area perpendicular to the settling direction. It is also confirmed by the monotonous increasing settling velocity. Overall, the simulation results matched well with the experimental one. Small deviations can be found at the beginning stage at $Re=0.57$, it can be explained by the fact that the initial releasing orientation of the metaball is not strictly controllable. Thus, the particle needs to adjust to the maximum projection area at the beginning stage. At low $Re$, the fluid velocity field surrounding the metaball is similar with sphere one as shown in Fig.~\ref{fig:a_meta_snapshot}, which is consistent with previous studies~\cite{zhang2016lattice}. The time evolution of hydrodynamic force acted on the metaball is plotted in Fig.~\ref{fig:meta_fh} for $Re$=8.74. It is clear that the sharp interface coupling scheme indeed introduced observable fluctuations in the hydrodynamic force, but the fluctuations are still within reasonable range and doesn't significantly affect the overall accuracy in terms of particle motions. Table 1 shows the averaged computational time per step for different functions without parallelization, it is found that LBM is the most time consuming part when the number of particles is small.

\begin{table}
\centering 
\begin{tabular}{ |c|c|c|c|c| }
\hline
DEM step & LBM collision & LBM stream  & IBB boundary condition\\
\hline
$6.65 \times 10^{-6}$ \si{s} & 1.13 \si{s} & 0.89 \si{s}  & $9.46 \times 10^{-3}$ \si{s}\\
\hline
\end{tabular}
\caption{\label{tab} Averaged computational time per step for difference functions without parallelization.}
\end{table}

\subsection{Instability issues for multiple particle simulations}
\begin{figure}
\centering     %%% not \center
\subfigure{\includegraphics[width=65mm]{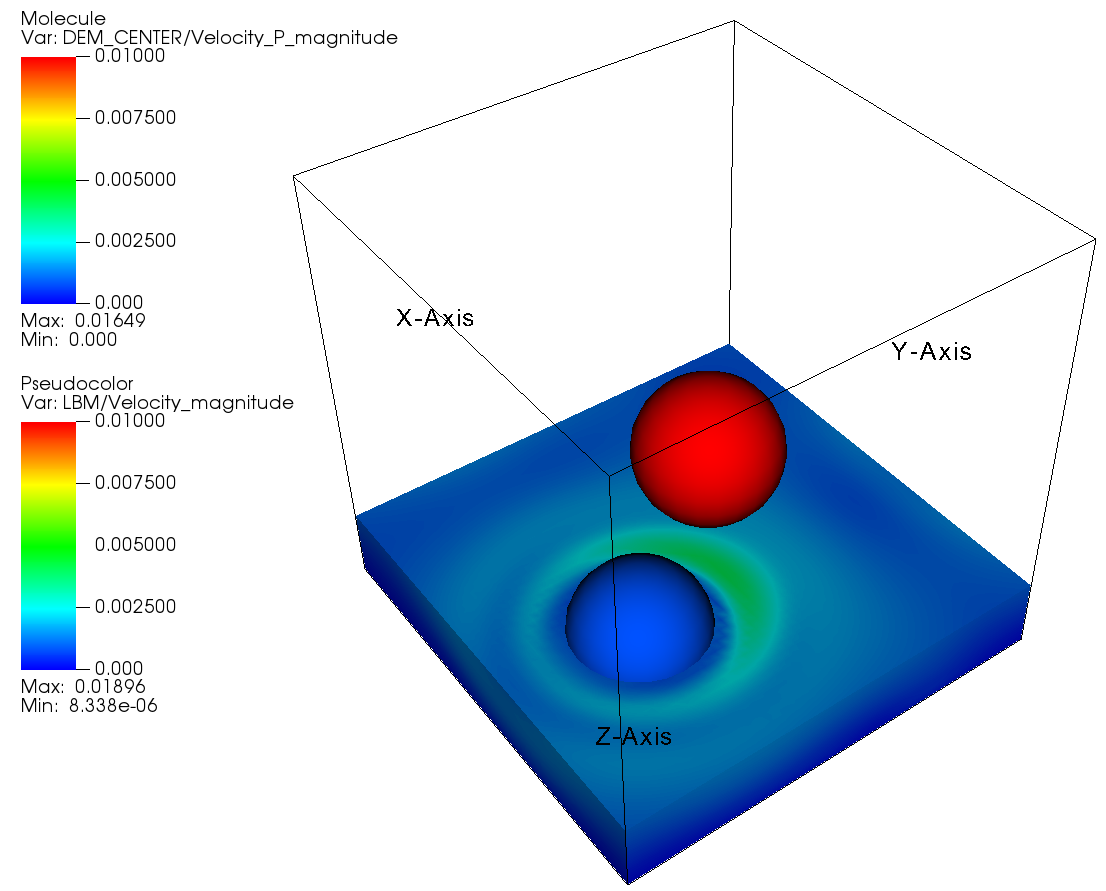}}
\subfigure{\includegraphics[width=65mm]{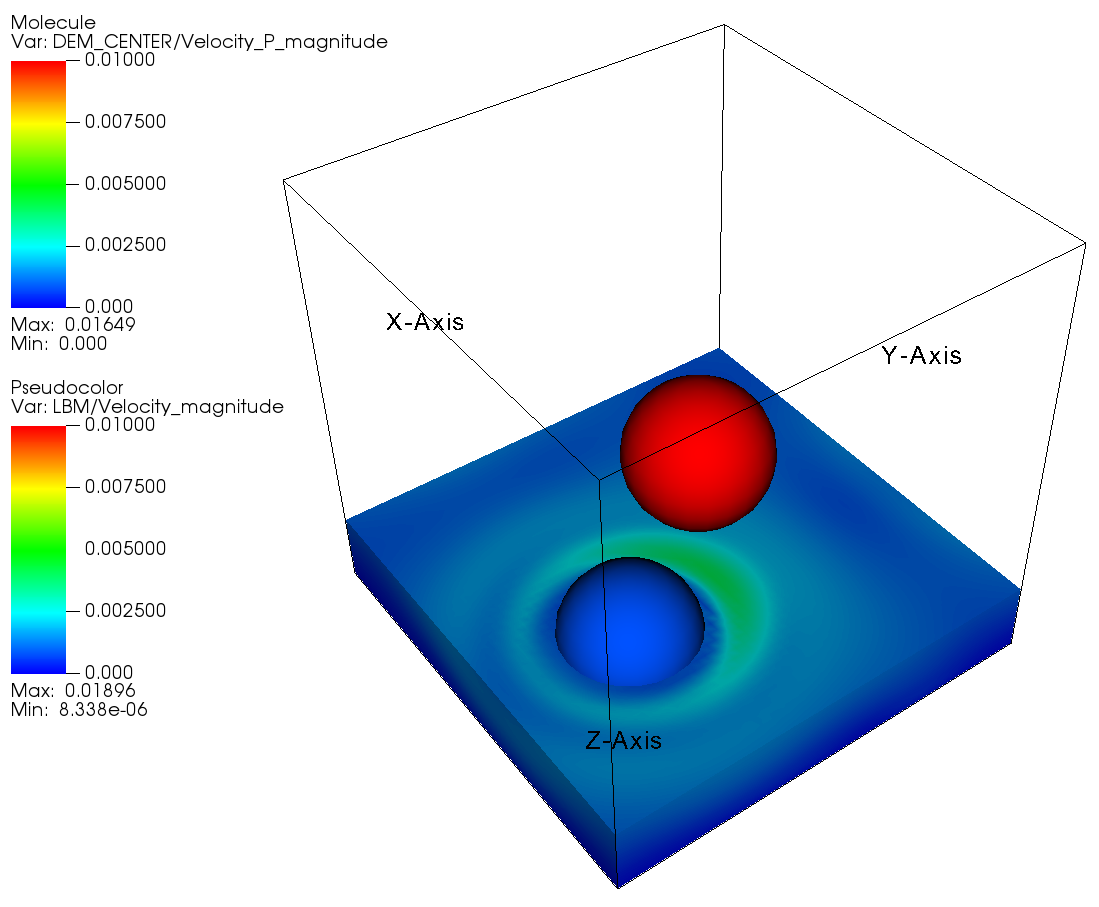}}
\subfigure{\includegraphics[width=65mm]{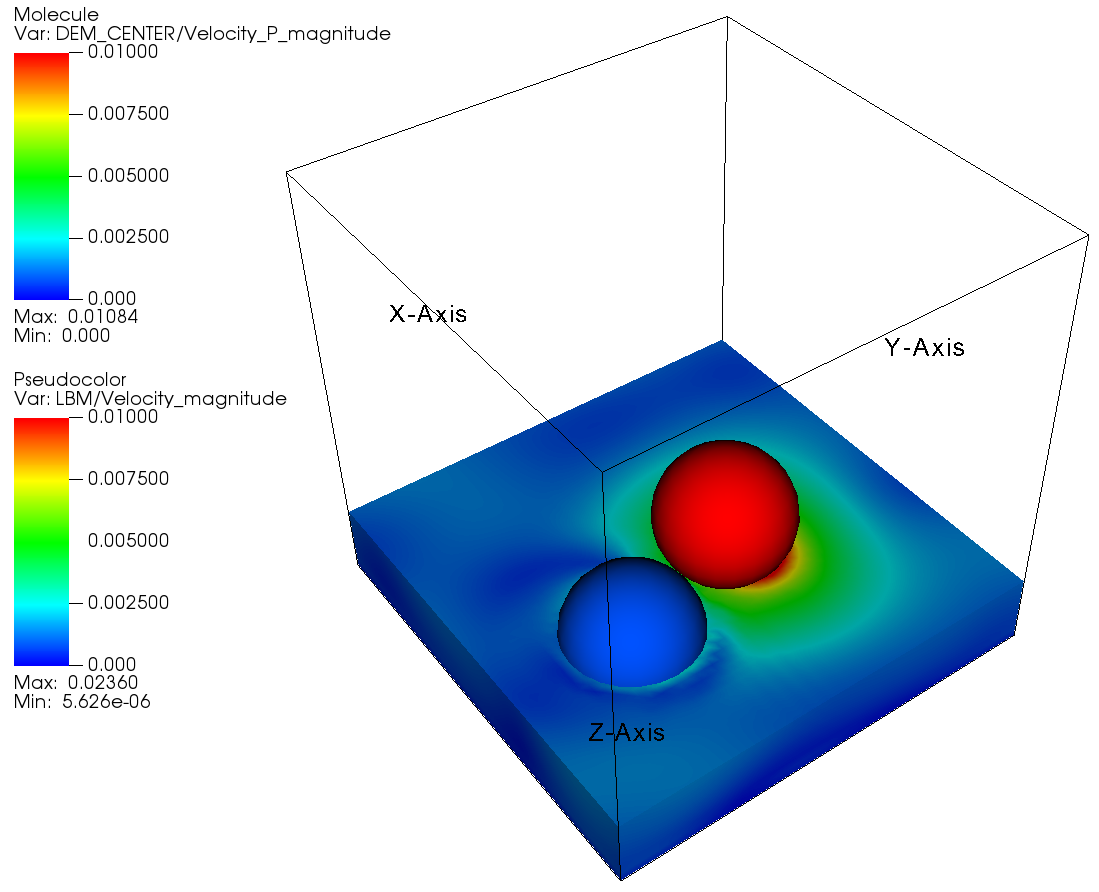}}
\subfigure{\includegraphics[width=65mm]{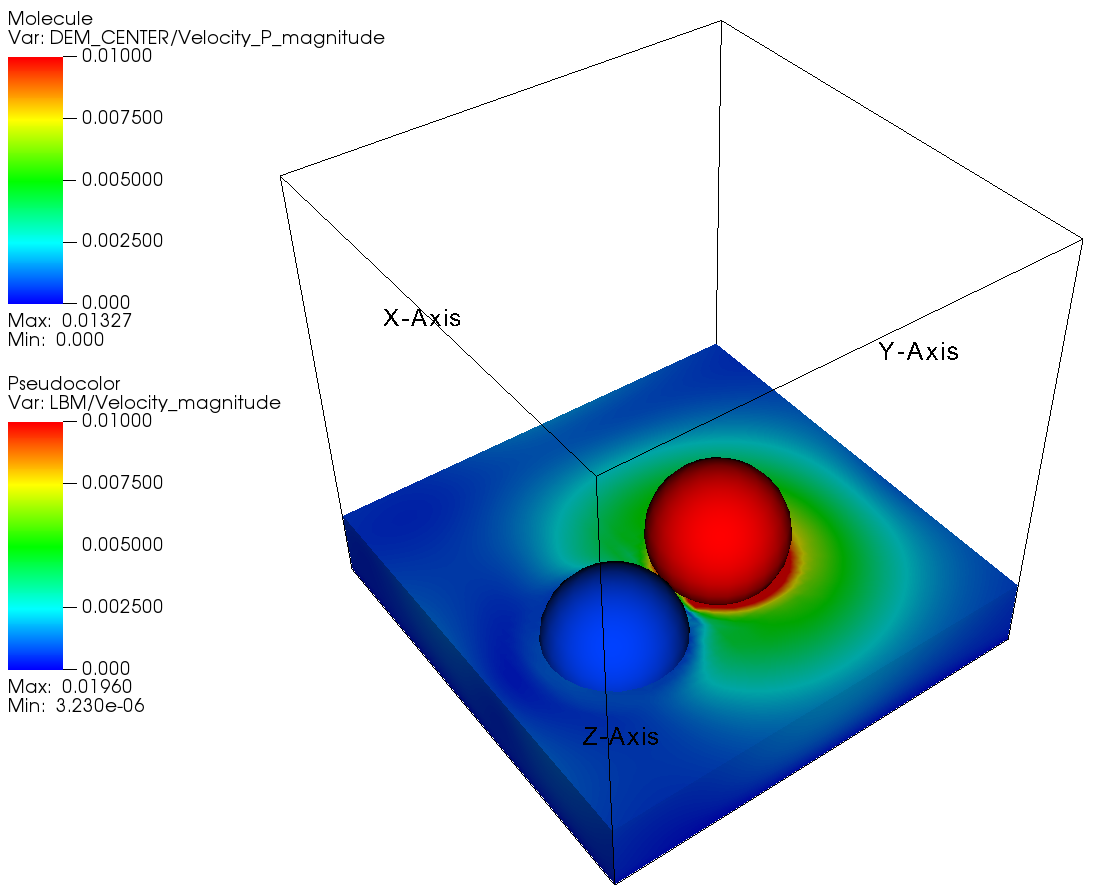}}
\subfigure{\includegraphics[width=65mm]{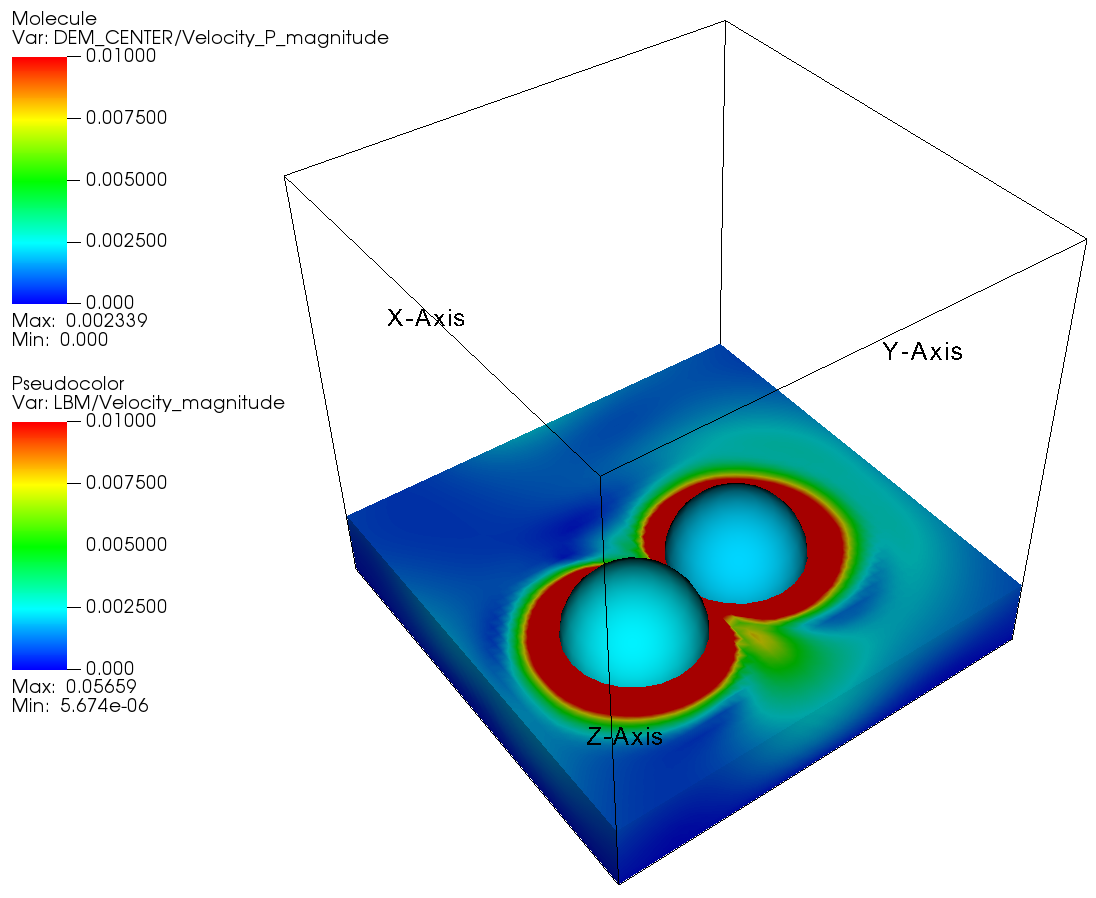}}
\subfigure{\includegraphics[width=65mm]{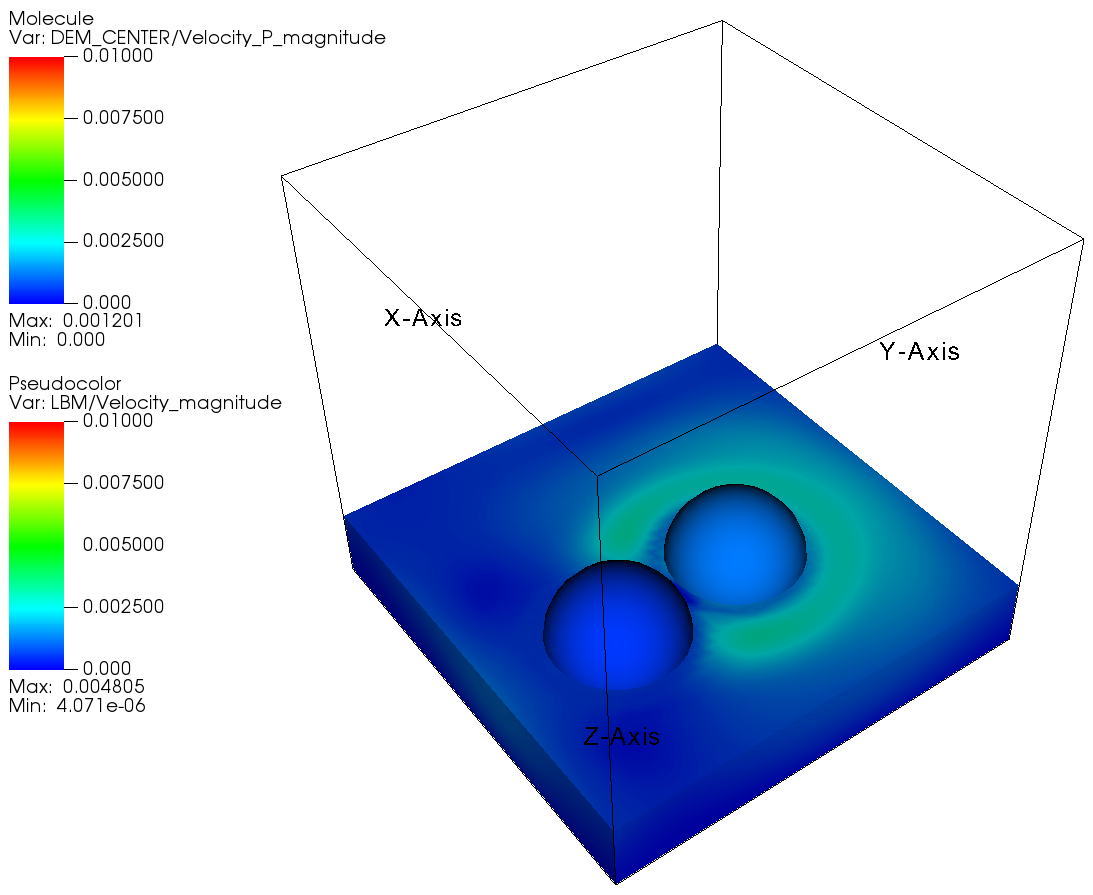}}
\subfigure{\includegraphics[width=65mm]{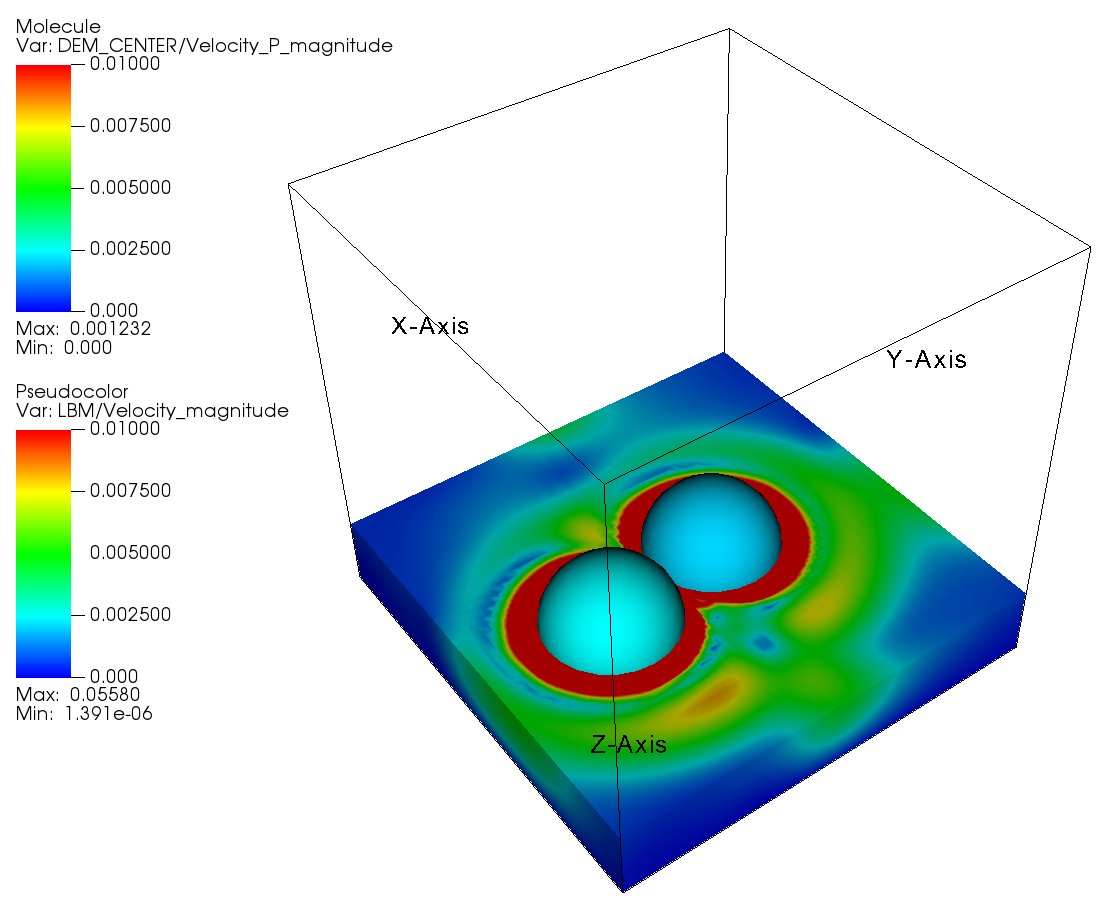}}
\subfigure{\includegraphics[width=65mm]{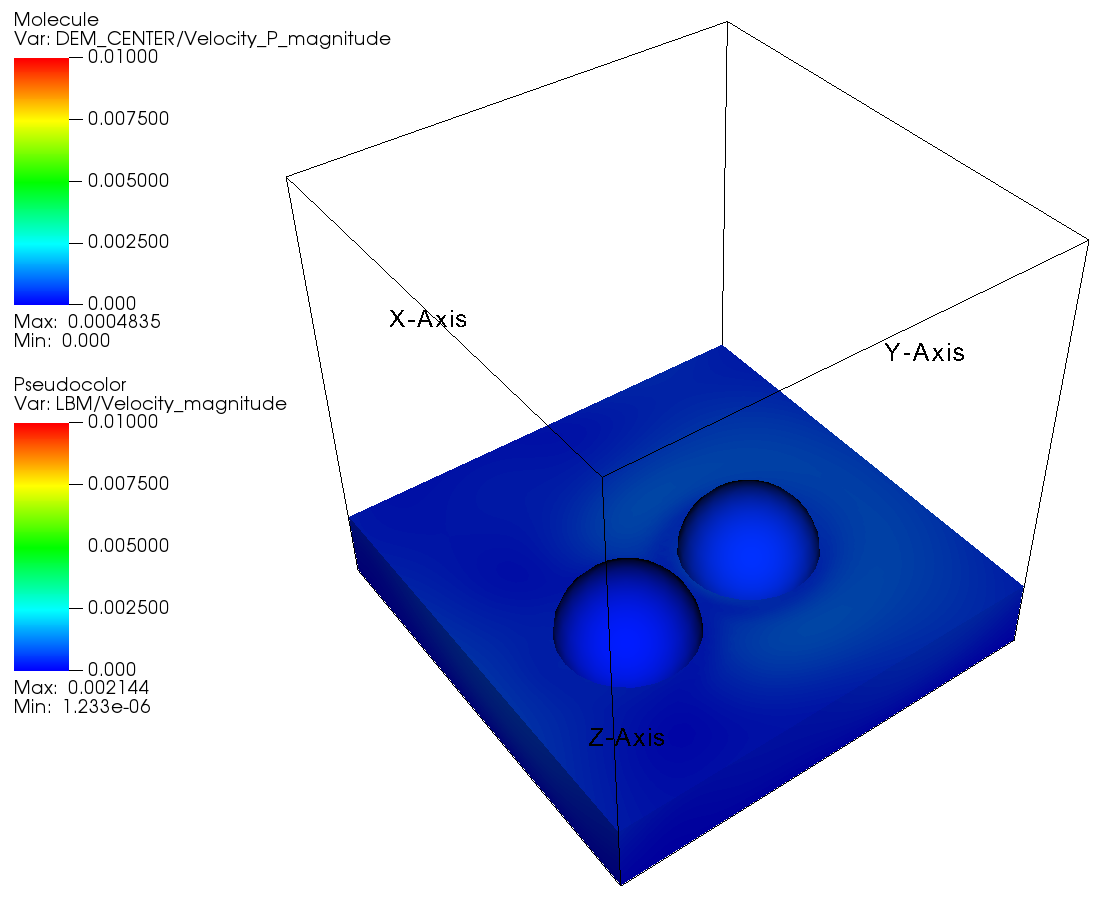}}
\caption{Snapshot of the metaball settling simulation for $Re=8.74$ at 0.3, 0.5, 0.7 and 1 s, colour indicates fluid velocity magnitude.}
\label{fig:issue}
\end{figure}

It is well known that sharp interface boundary conditions suffer from spurious hydrodynamic force oscillations~\cite{seo2011sharp}. This issue can be even more profound for multiple particle simulations due to the lack of enough resolution between particles. The oscillating forces and torques can cause numerical instabilities, even crash simulations. Our analysis in the previous section shows that the lack of isotropic on forcing point distribution is an important source of nonphysical oscillations. To illustrate this problem, the settling of two spheres is conducted with the same parameters in section~\ref{sec:settling_sphere}. The time series of particle positions and fluid velocity field are shown in Fig.~\ref{fig:issue}, where the left panel shows simulations without treatment and the right panel with treatments described in section~\ref{sec:treatment} (Eq.~\ref{eq:fh_s_issue} and Eq.~\ref{eq:th_s_issue}). The simulations are identical before two particles reach the bottom. However, particles produce nonphysical spins without treatments and the numerically introduced energy cannot be dissipated (see the last two figures in the left panel of Fig.~\ref{fig:issue}). On the contrary, both particle and fluid velocities reach zero if the isotropic of forcing point distribution is restored. In fact, the simulation shown in the left panel becomes unstable if the simulation continues, where no stability issue is found with treatments.

\section{Numerical examples}\label{sec:examples}
\begin{figure}[t]
\begin{centering}
\includegraphics[width=0.5\linewidth]{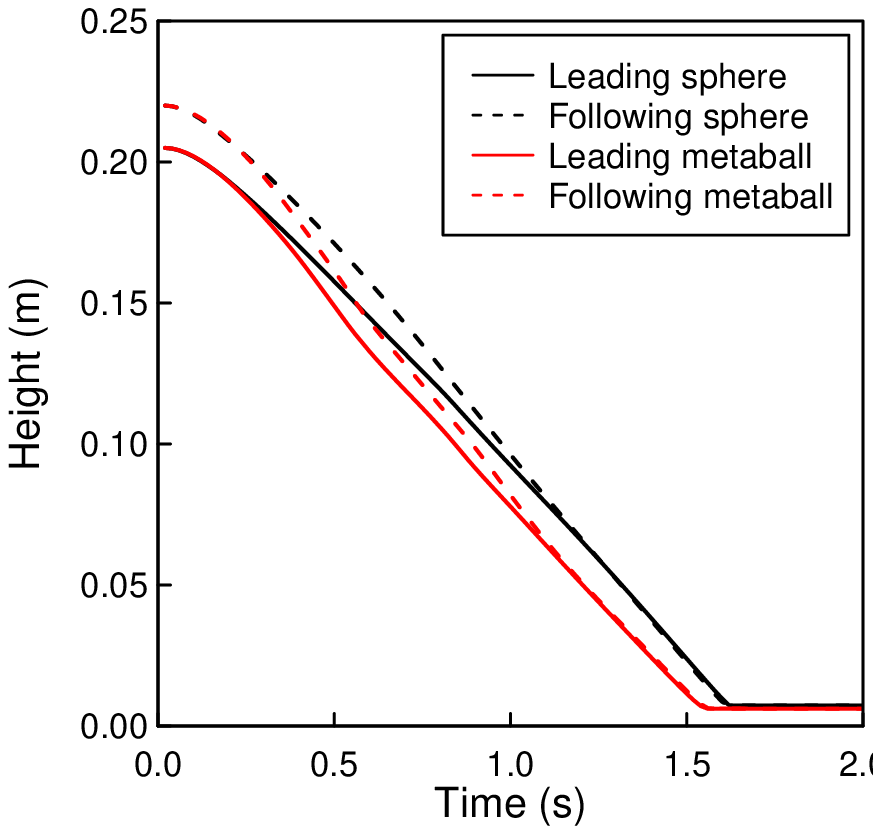}
\caption{Time series of height of particle centre. Solid line and dotted line represent represent the leading particle and t he following particle, respectively. Black for sphere and red for metaball.}
\label{fig:dkt_h}
\end{centering}
\end{figure}

\begin{figure}[t]
\begin{centering}
\includegraphics[width=0.5\linewidth]{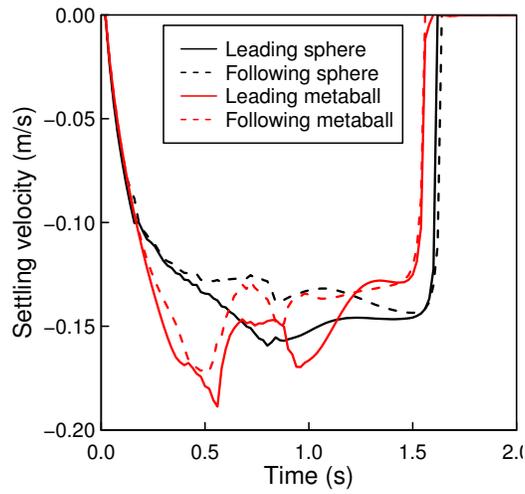}
\caption{Time series of particle's vertical velocity. Solid line and dotted line represent represent the leading particle and t he following particle, respectively. Black for sphere and red for metaball.}
\label{fig:dkt_v}
\end{centering}
\end{figure}

\begin{figure}[t]
\begin{centering}
\includegraphics[width=0.5\linewidth]{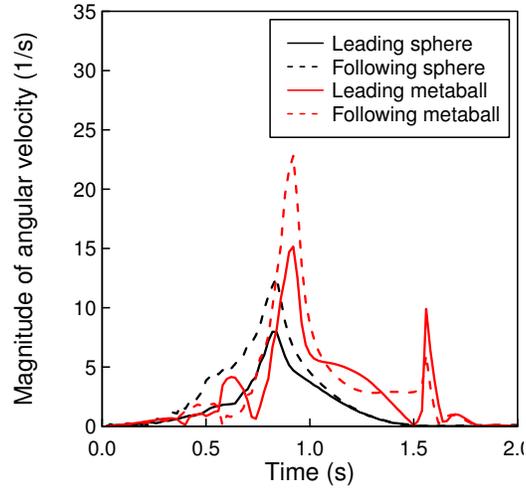}
\caption{Time series of angular velocity magnitude. Solid line and dotted line represent represent the leading particle and t he following particle, respectively. Black for sphere and red for metaball.}
\label{fig:dkt_w}
\end{centering}
\end{figure}

\begin{figure}
\centering     %%% not \center
\subfigure{\includegraphics[width=30mm]{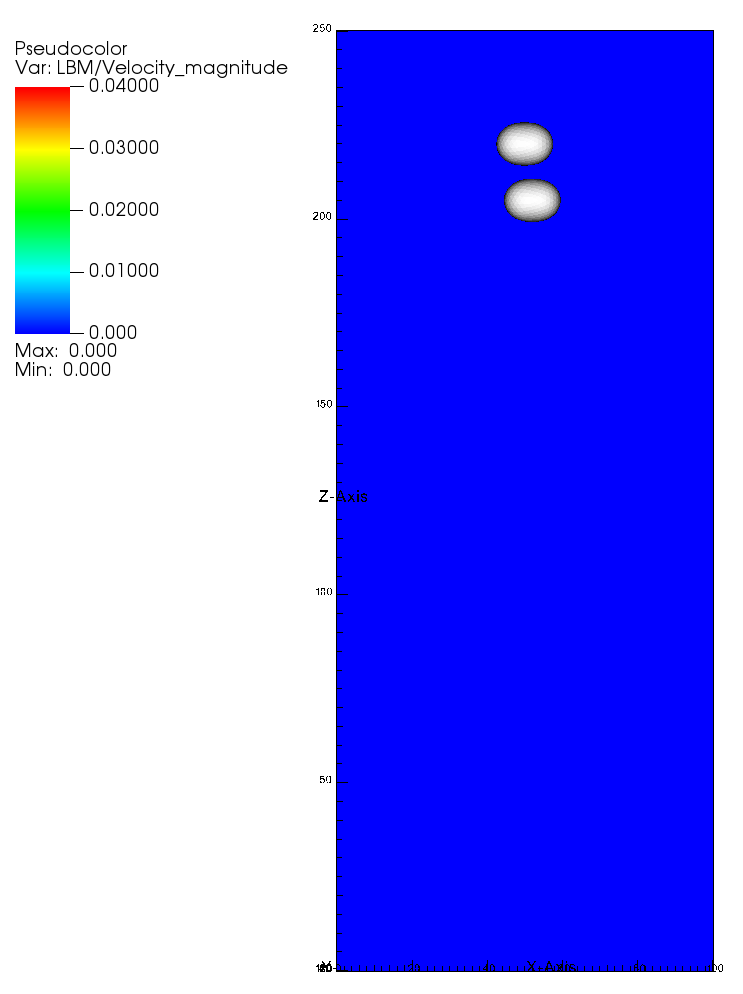}}
\subfigure{\includegraphics[width=30mm]{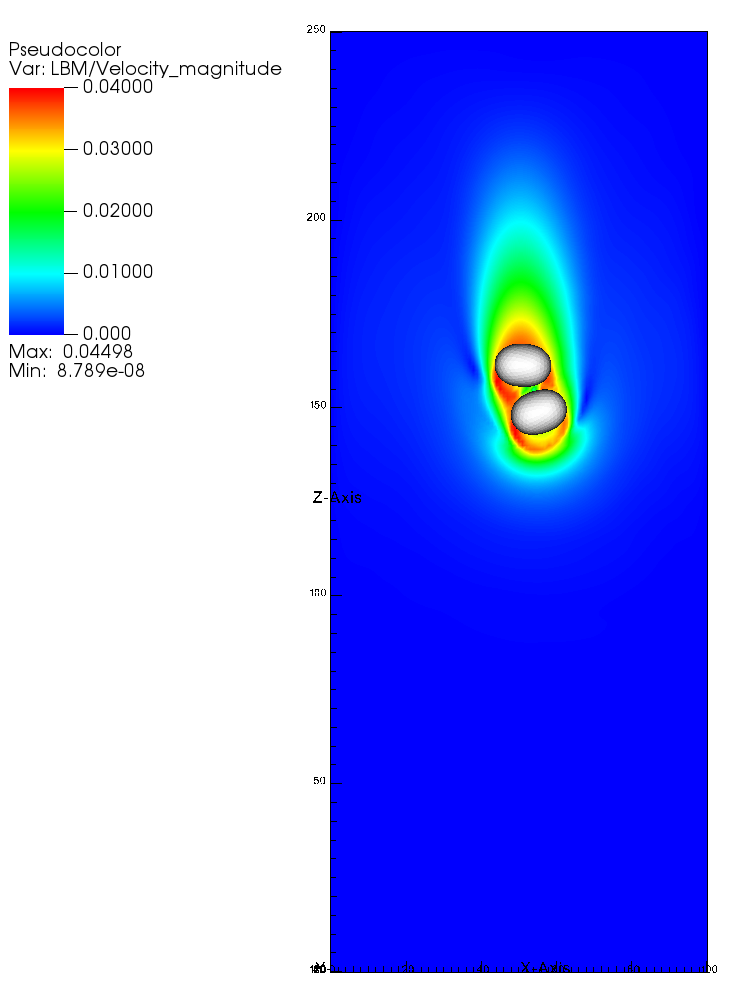}}
%\subfigure{\includegraphics[width=30mm]{dkt_35.png}}
\subfigure{\includegraphics[width=30mm]{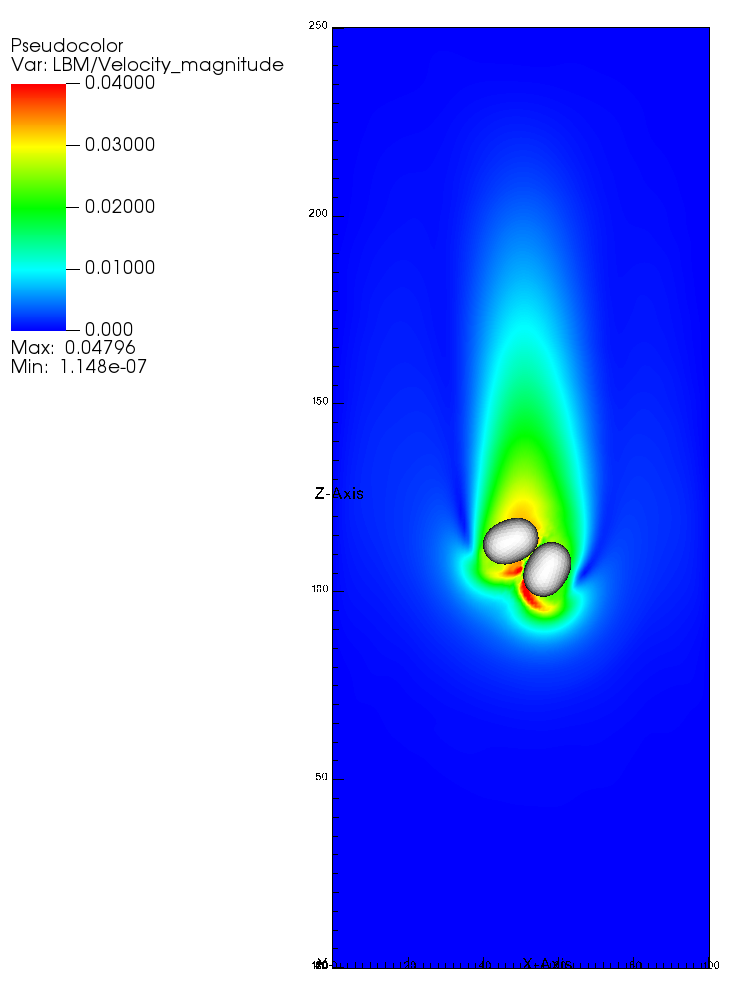}}
\subfigure{\includegraphics[width=30mm]{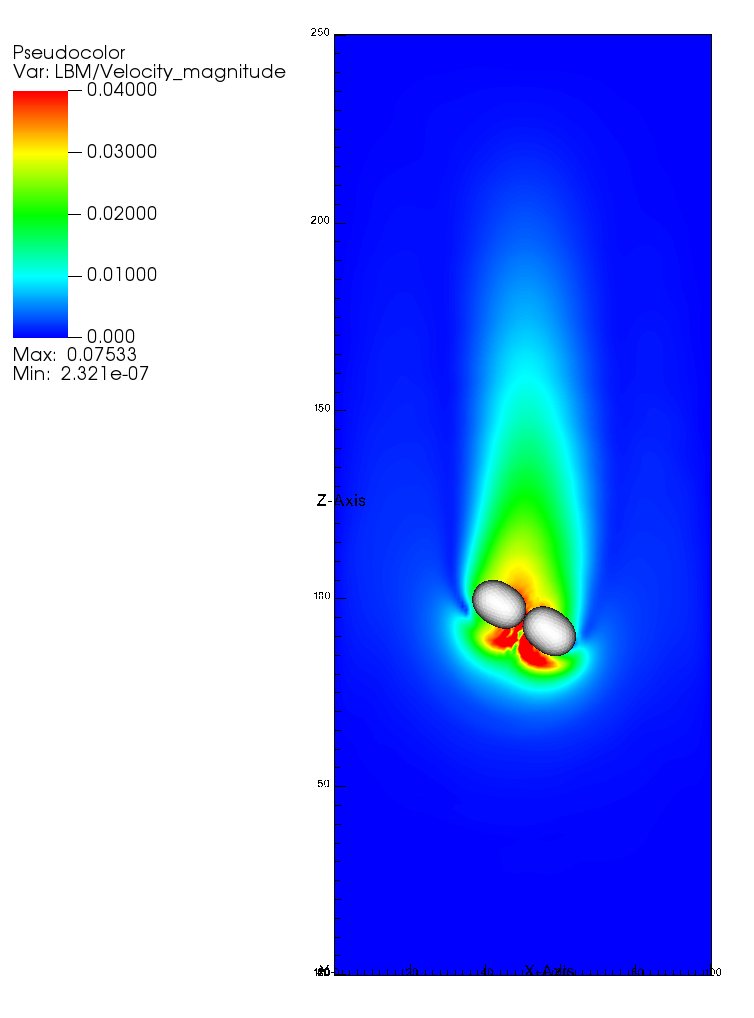}}
\subfigure{\includegraphics[width=30mm]{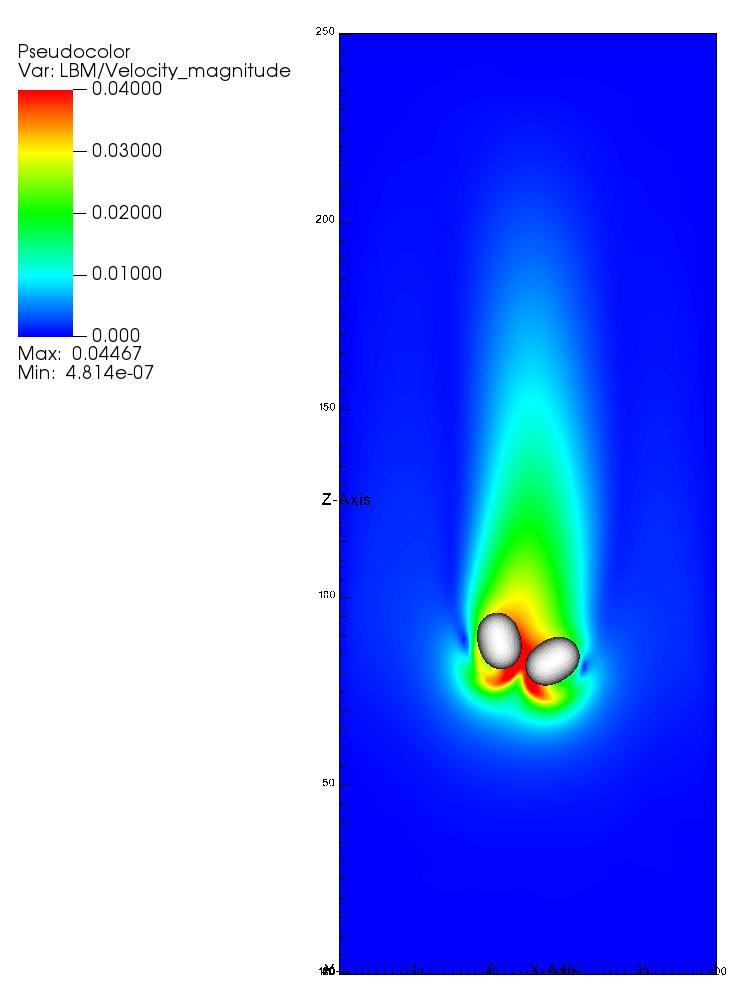}}
\subfigure{\includegraphics[width=30mm]{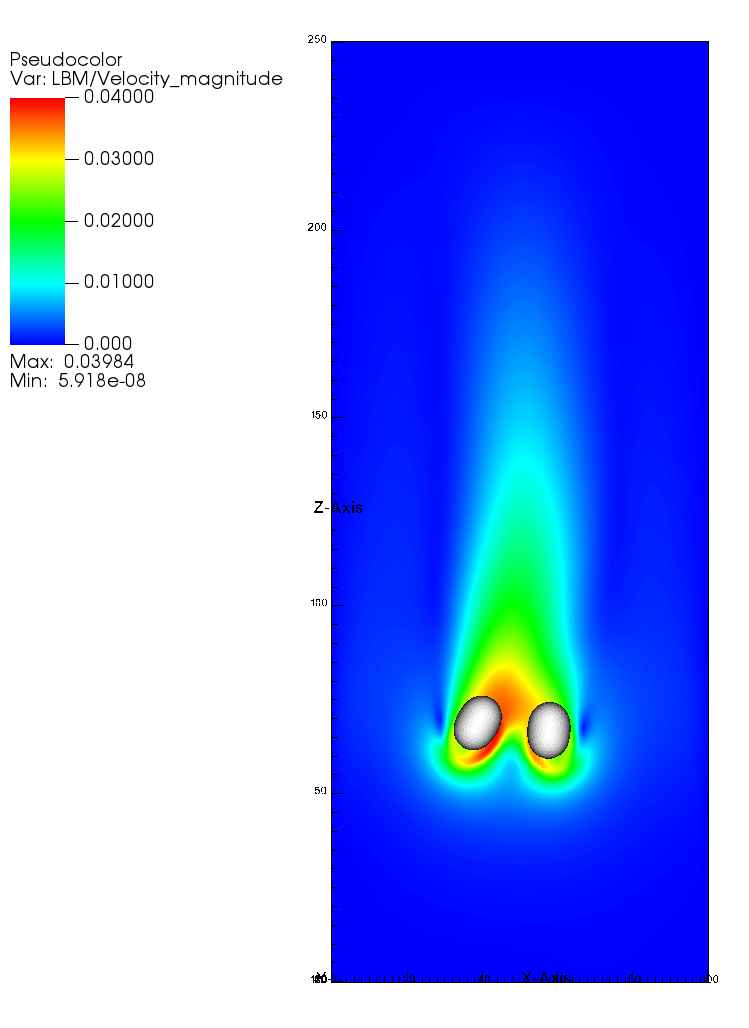}}
\subfigure{\includegraphics[width=30mm]{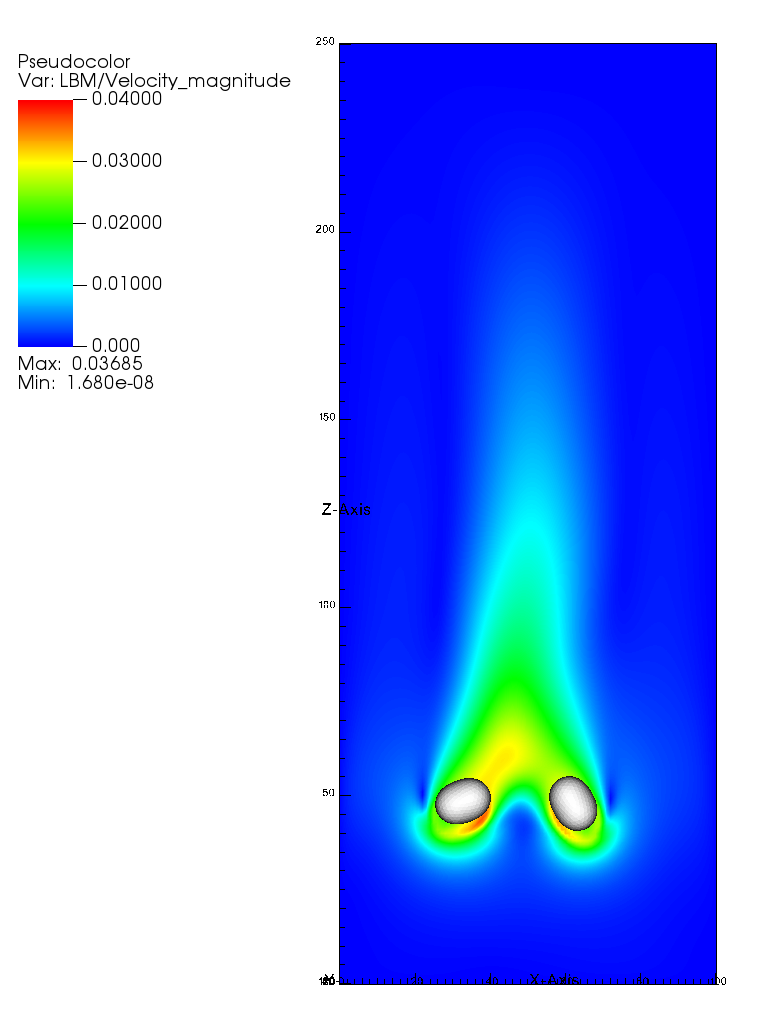}}
\subfigure{\includegraphics[width=30mm]{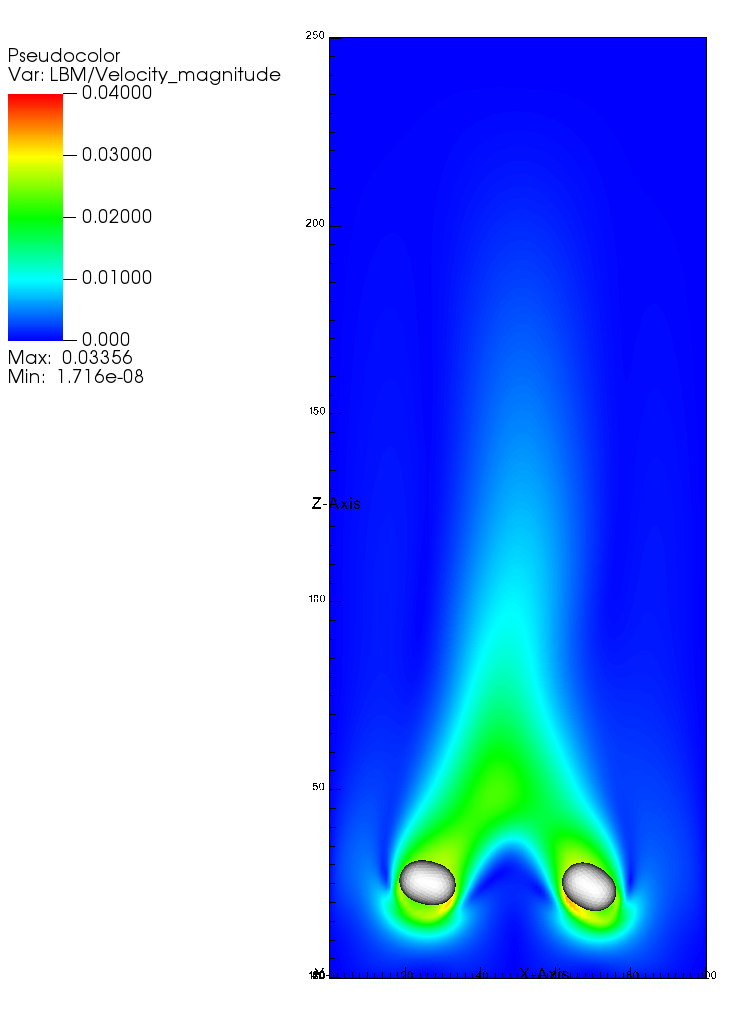}}
%\subfigure{\includegraphics[width=30mm]{dkt_80.png}}
\caption{Snapshot of the two metaball settling simulation at 0, 0.5, 0.8, 0.9, 0.96, 1.08, 1.22 and 1.4 s, colour indicates fluid velocity magnitude.}
\label{fig:dkt}
\end{figure}

%\subsection{Settling of two metaballs}\label{sec:dkt}
In this section, a simulation of two settling metaballs is conducted in a closed box with dimension of $0.15 \times 0.15 \times 0.25$ \si{m}. Same metaballs as in section~\ref{sec:settling_metaball} are used but with density $1200$ \si{kg/m^3}. The fluid density and viscosity are $927$ \si{kg/m^3} and $7.55 \times 10^{-5}$ \si{m^2/s}. Two metaballs are placed at $0.22$ and $0.205$ \si{m} from the bottom initially, and the lower metaball is placed $0.002$ \si{m} off from the centreline. To highlight the importance of capturing particle shapes, the same simulation with spheres is also conducted, where the spheres have the same volume as the metaballs. It is well known that the settling of two spheres under gravity has complex dynamics often referred to as "drafting, kissing and tumbling" (DKT), which was first numerical studied by Feng and Michaelides~\cite{feng2004immersed}. In DKT, the following particle will catch up with the leading particle due to the drag reduction from the leading particle's wake. A similar trend is found for the settling of two metaballs in this simulation. Fig.~\ref{fig:dkt_h},~\ref{fig:dkt_v} and~\ref{fig:dkt_w} show time series of height of particle centre, particle vertical velocity and angular velocity magnitude for both metaballs and spheres. The drag coefficient for non-spherical particles is generally larger than for volume equivalent spheres~\cite{zhang2016lattice}. Surprisingly, metaballs reach the bottom before spheres, although the initial projection area of metaball is larger than the sphere. Fig.~\ref{fig:dkt} reveals detailed flow patterns around metaballs as well as particle positions and orientations. It is clear that the particle shape induced rotations have significant influences on particle dynamics: rotations can reduce the projection area and result in a lower drag force. The effects of variable projection area can also be observed by the time series of settling and angular velocities and, where the fluctuation of metaball settling and angular velocities is considerably larger than those of sphere's (Fig.~\ref{fig:dkt_v}).

\begin{figure}
\centering     %%% not \center
\subfigure{\includegraphics[width=40mm]{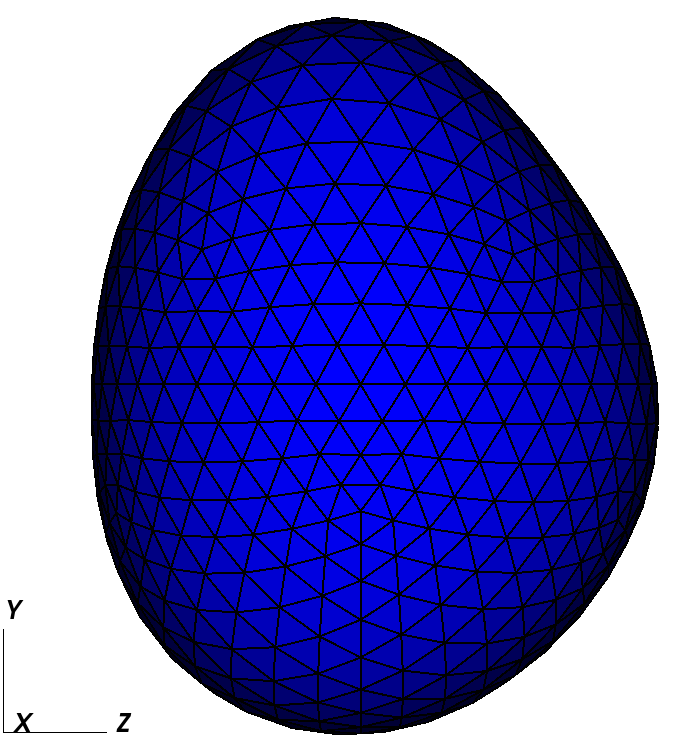}}
\subfigure{\includegraphics[width=40mm]{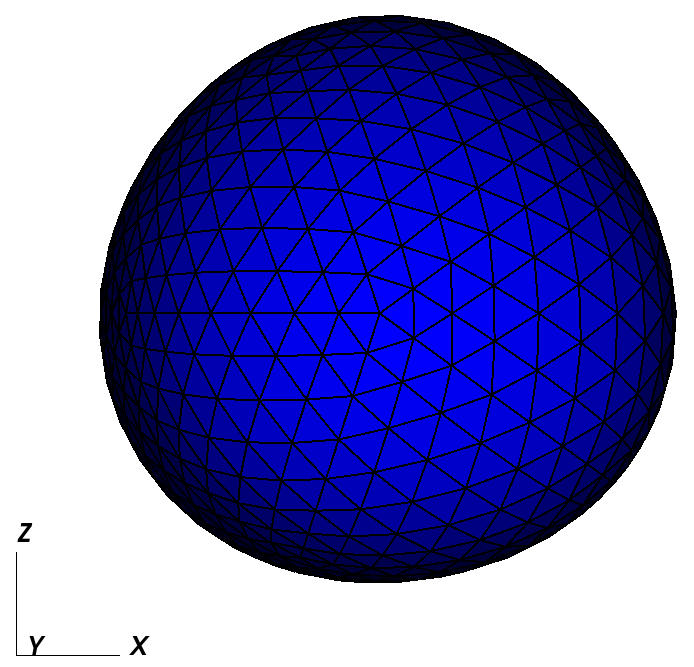}}
\subfigure{\includegraphics[width=40mm]{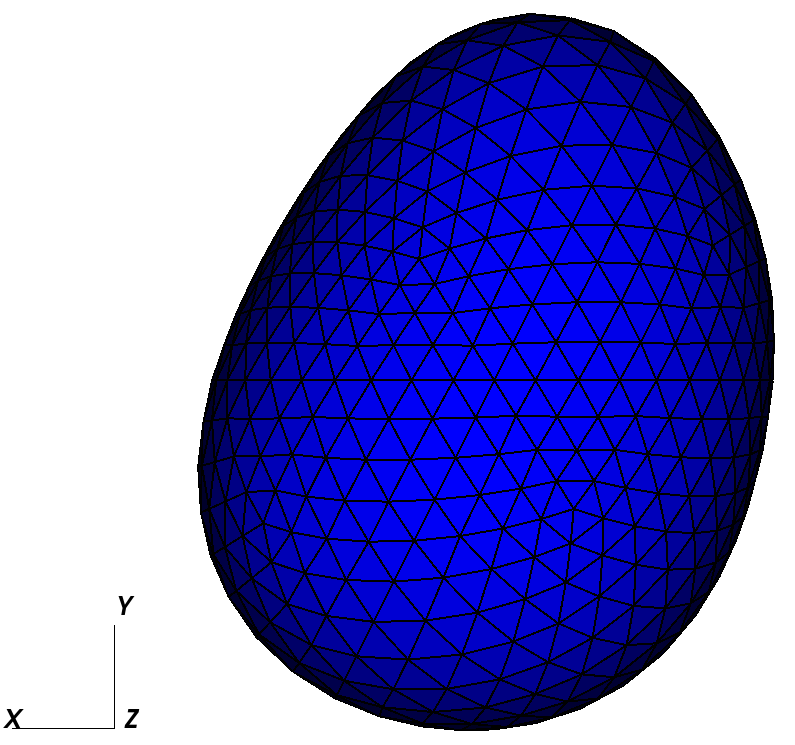}}
\caption{Side views of the irregularly shaped metaball.}
\label{fig:s30_p}
\end{figure}

\begin{figure}
\centering     %%% not \center
\subfigure{\includegraphics[width=40mm]{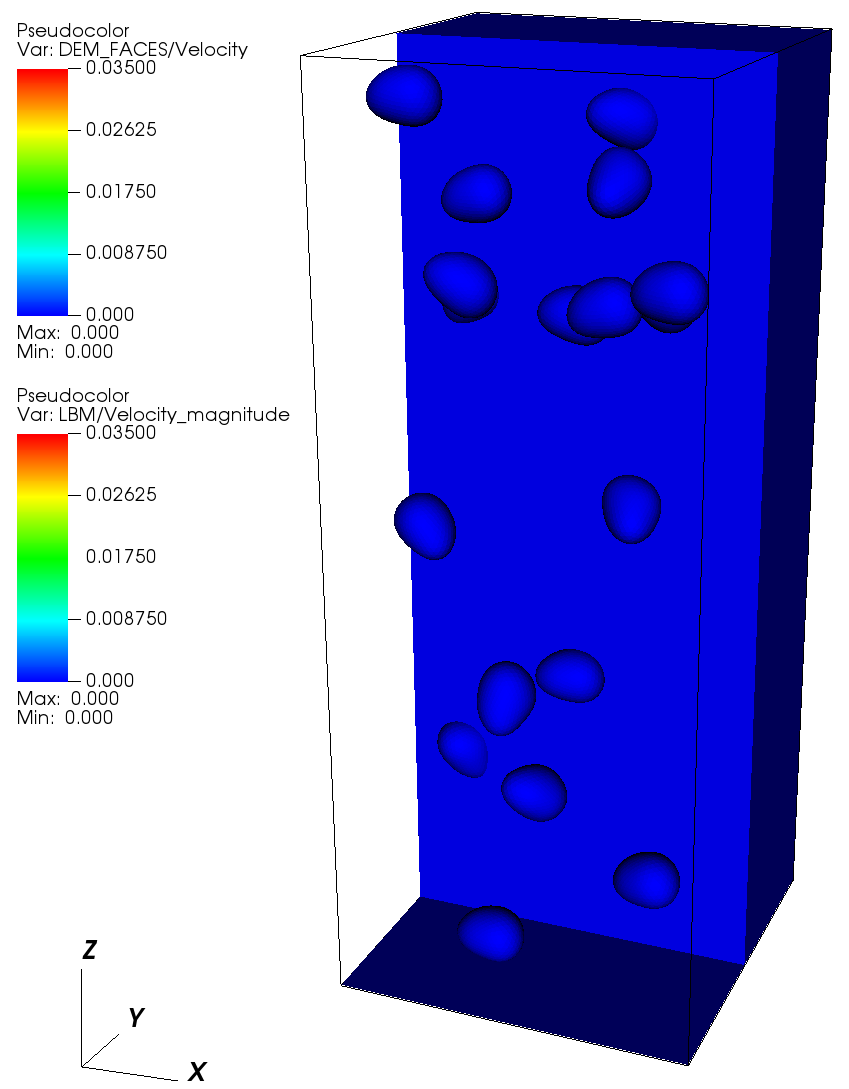}}
\subfigure{\includegraphics[width=40mm]{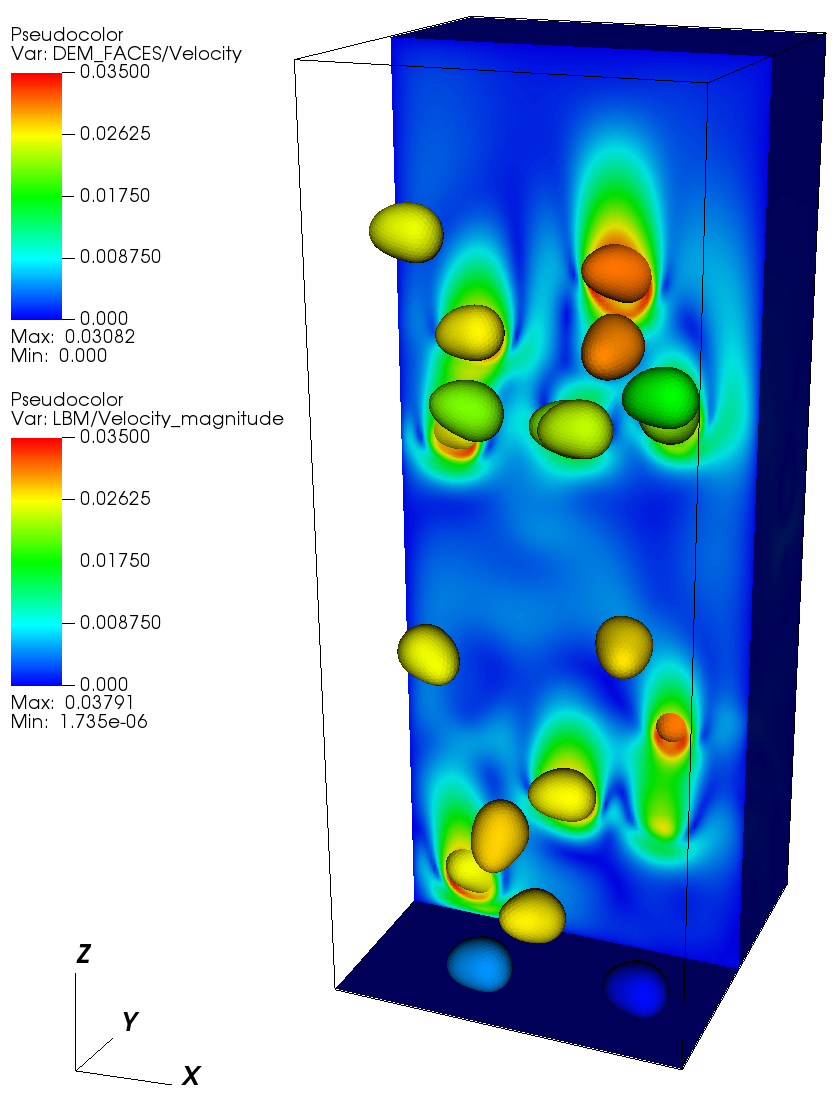}}
\subfigure{\includegraphics[width=40mm]{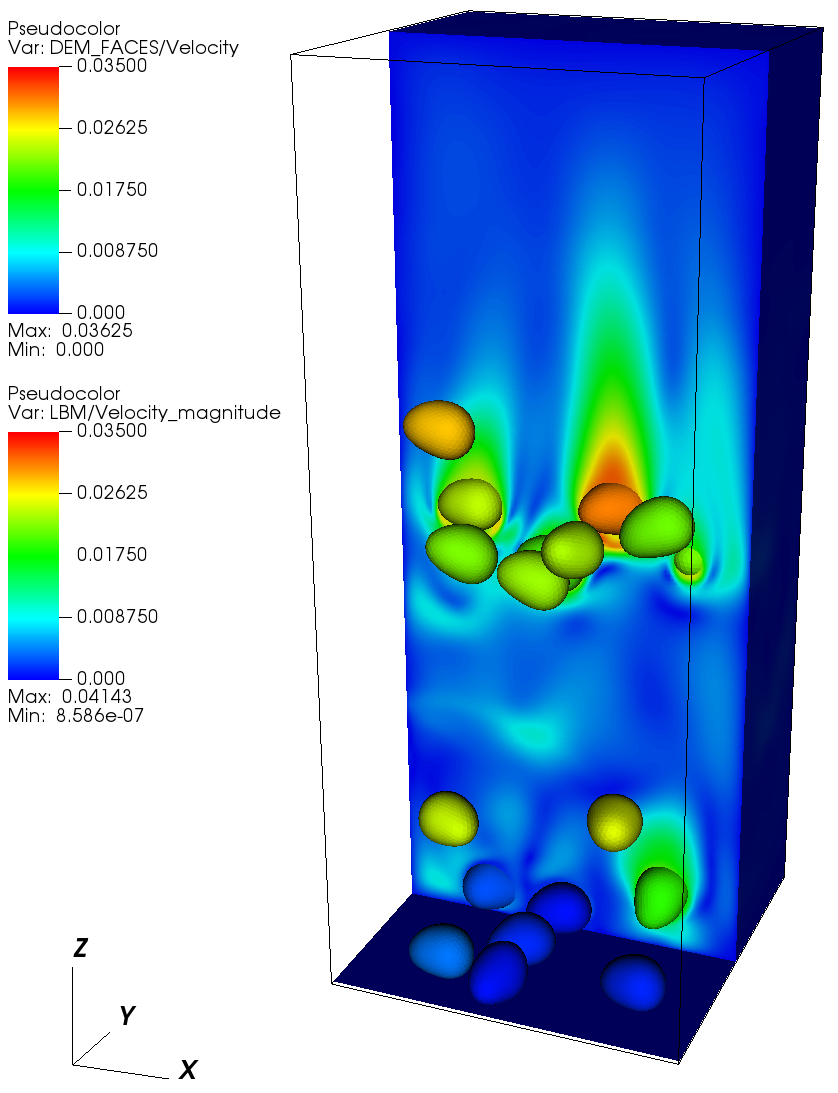}}
\subfigure{\includegraphics[width=40mm]{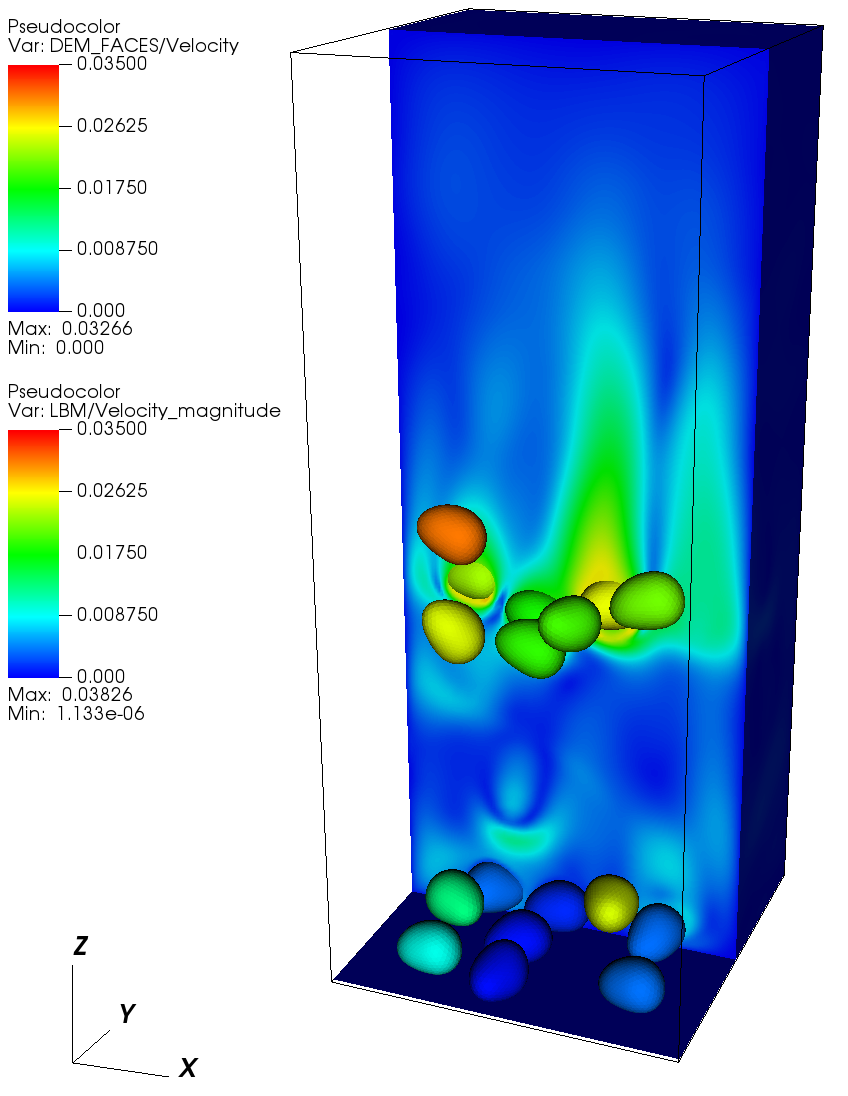}}
\subfigure{\includegraphics[width=40mm]{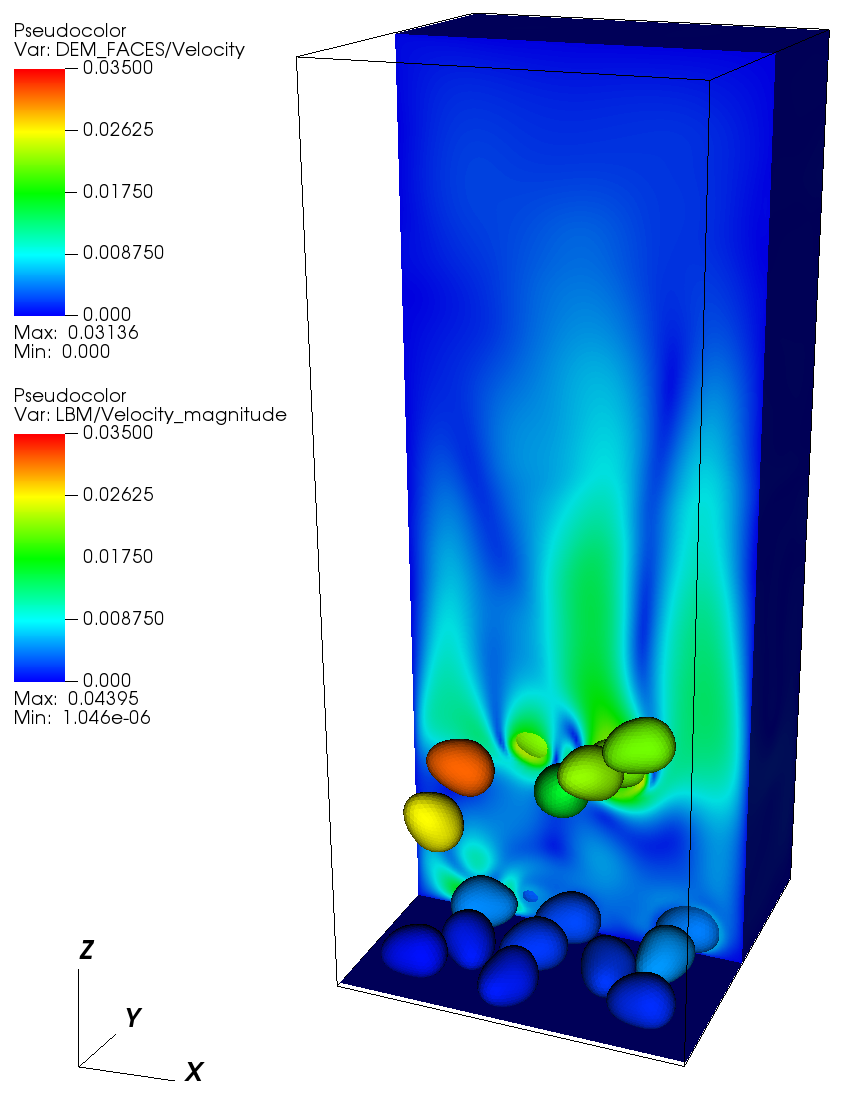}}
\subfigure{\includegraphics[width=40mm]{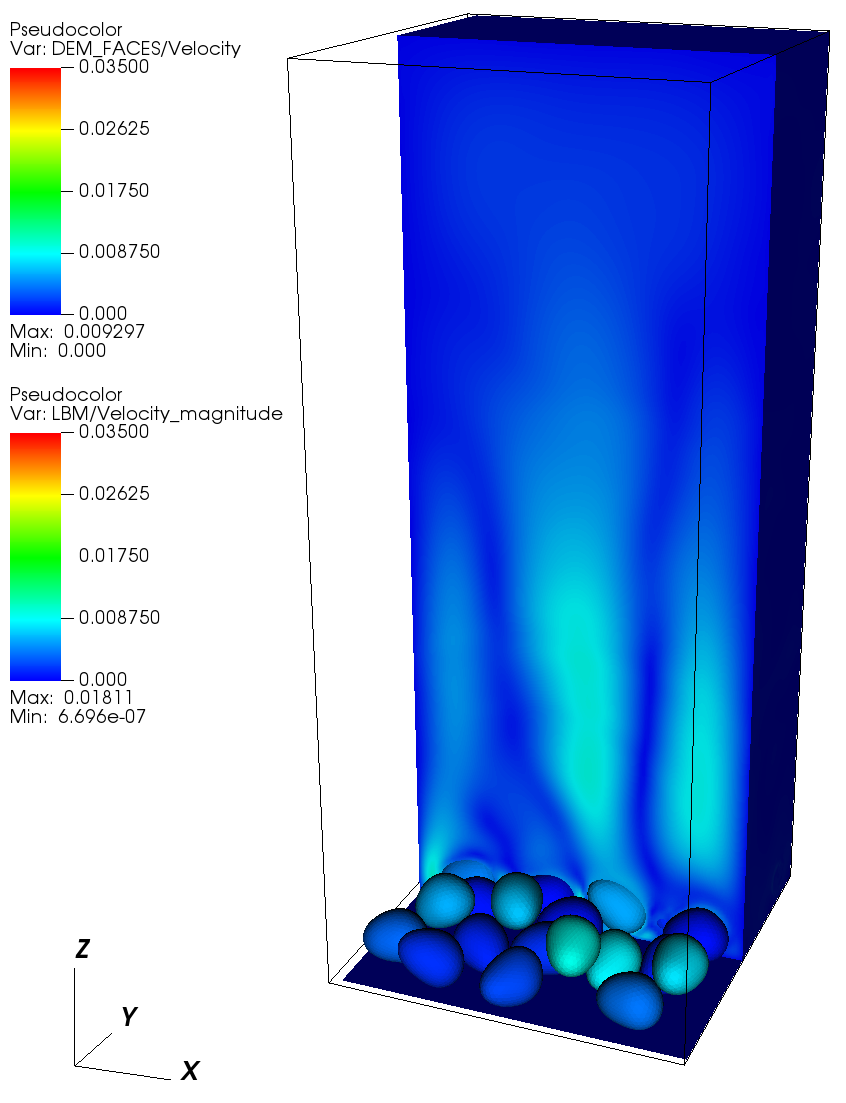}}
\caption{Snapshot of the 30 metaball settling simulation at 0, 0.4, 0.8, 1, 1.4 and 2 s, colour indicates fluid velocity magnitude.}
\label{fig:s30}
\end{figure}

%\subsection{Settling of many metaballs}
The last example is the settling of 30 metaballs with complex shapes. The side views of a randomly generated particle shape are shown in Fig.~\ref{fig:s30_p}, it is clear that the shape is non-isotropic. Same parameters are used as in the previous section but with periodic boundary conditions applied to horizontal directions and the particle volume is $2.186 \times 10^{-6}$ \si{m^3}. 30 metaballs are randomly placed within the domain with zero initial velocity and start to settle under gravity. Fig.~\ref{fig:s30} shows the time evolution of the fluid-particle systems with detailed flow structures, which demonstrated the capability of the proposed model in simulating fluid-particle systems with complex particle shapes.

\section{Concluding remarks}\label{sec:conclusion}
In this paper, we proposed a coupled metaball DEM-LBM model to simulate fluid-particle, particle-particle interactions with complex particle shapes. By introducing a proper sharp interface coupling scheme, the efficiency of LBM in solving flows and the capability of metaball DEM in handling non-spherical particles are integrated. To preserve the high accuracy of sharp interface moving boundary conditions, its numerical instability issues are addressed by a local refilling algorithm and the re-evaluation of hydrodynamic forces from solid nodes. Implementations of metaball DEM, LBM, and the coupling scheme are presented in detail.

The proposed model is first validated by comparing settling velocities of a sphere (with metaball representation) in the fluid under various $Re$ with experimental results, good agreements are observed for the settling velocities. By simulating the settling of a non-spherical metaball and comparing with our experimental results, the model shows its capability in accurately handling fluid-particle interactions with complex particle shapes. The treatments of instability issues and their effects are illustrated by multiple particle simulations, which suggests that the proposed coupling scheme can efficiently suppress the non-physical spin when two particles are close to each other.

To demonstrate the capability of the model for fluid-particle systems with complex particle shapes, the classic DKT phenomenon is reproduced with non-spherical shapes. It is found that shapes have significant effects on particle dynamics, it is essential to accurately capture particle shape. The model is then applied to simulate the settling of 30 metaballs which clearly shows that the model can handle complex particle geometries.

In conclusion, the presented results demonstrate the potential of the metaball DEM-LBM model as a powerful numerical tool for simulating a wide range of fluid-particle systems, particularly for non-spherical particles which can be found in many engineering and science disciplines.

\section*{Acknowledgement}\label{sec:Acknowledgement}
We gratefully acknowledge the funding from the Zhejiang Provincial Key Research and Development Program (2021C02048), Natural Science Foundation of Zhejiang Province, China (LHZ21E090002) and the National Natural Science Foundation of China (12172305).


%% References


\bibliographystyle{elsarticle-num}

\bibliography{mybib.bib}






% \end{thebibliography}

\end{document}